\documentclass[aps,prd,onecolumn,eqsecnum,amsmath,nofootinbib,preprintnumbers]{revtex4}%

\usepackage{color,graphicx,float,subfigure}
\usepackage{amsfonts,amssymb,theorem,mathrsfs,times}
\usepackage{bm}
\usepackage{mathtools}
\usepackage{amsfonts,amssymb,theorem,mathrsfs}
\usepackage{dsfont}
\usepackage{setspace}

{\theorembodyfont{\upshape}
}
{\theorembodyfont{\upshape}
}
{\theorembodyfont{\upshape}
}
{\theorembodyfont{\upshape}
}
{\theorembodyfont{\upshape}
}
{\theorembodyfont{\upshape}
}

\newcommand{\dalm}{\kern1pt\vbox{\hrule height 0.9pt\hbox{\vrule width
0.9pt\hskip 2.5pt\vbox{\vskip 5.5pt}\hskip 3pt\vrule width
0.3pt}\hrule height 0.3pt}\kern1pt}

\begin{document}
\preprint{\hfill {\small {ICTS-USTC/PCFT-21-02}}}
\title{Hyperbolicity and Causality of Einstein-Gauss-Bonnet Gravity  in Warped Product Spacetimes   }

%

\author{ Li-Ming Cao$^{a\, ,b}$\footnote{e-mail
address: caolm@ustc.edu.cn}}

\author{ Liang-Bi Wu$^b$\footnote{e-mail
address: liangbi@mail.ustc.edu.cn}}

\affiliation{$^a$Peng Huanwu Center for Fundamental Theory, Hefei, Anhui 230026, China}

\affiliation{${}^b$
Interdisciplinary Center for Theoretical Study and Department of Modern Physics,\\
University of Science and Technology of China, Hefei, Anhui 230026,
China}


\date{\today}

\begin{abstract}
In Einstein-Gauss-Bonnet gravity, for a group of warped product spacetimes, we get a generalized master equation for  the  perturbation of tensor type.  We show that
the ``effective metric" or ``acoustic metric" for the tensor perturbation equation can be defined even without a static condition.  Since this master equation does not depend on the mode expansion,  the hyperbolicity and causality  
of the tensor perturbation equation can be investigated for every mode of the perturbation.  Based on the master equation, we study the hyperbolicity and causality for all relavent vacuum solutions of this theory. For each solution, we give the exact hyperbolic condition of the tensor perturbation equations.  Our approach can also applied to dynamical spacetimes, and
Vaidya spacetime have been investigated as an example.
\end{abstract}


\maketitle

 

\section{Introduction}
Lovelock theories are the most general diffeomorphism covariant theories only involving a metric tensor with second order equations of motion~\cite{Lovelock:1971yv}. In four dimension and generic values of the coupling constants, the theory reduces to the General Relativity with a cosmological constant~\cite{Lovelock:1972vz}. The equations of motion of such a theory in four dimensions are the Einstein equations. However, when the higher order terms of spacetime curvature exist, there will be a lot of properties which are different from the Einstein equations. For example, the gravitational propagation velocity may exceed the speed of light unlike the Einstein theories where graviton always travels slower than  light~\cite{ChoquetBruhat:1988dw}.

 Einstein-Gauss-Bonnet gravity, whose Lagrangian contains the quadratic term of spacetime curvature, as the lowest Lovelock theory,  is a simplest model to display the difference between the general Lovelock gravity theory and  Einstein gravity theory in higher dimensions. The theory is fascinating since it has been realized in the low-energy limit of heterotic string theory~\cite{Gross:1986iv, Gross:1986mw, Metsaev:1987bc, Metsaev:1987zx}. Moverover, this theory is studied in many aspects, such as black holes~\cite{Boulware:1985wk, Wheeler:1985nh, Cai:2001dz, Cai:2003gr} and AdS/CFT correspondence~\cite{Brigante:2008gz, Brigante:2007nu}.

It is well known that Einstein equations are second order quasi-linear equations, i.e., the coefficients of the second order partial derivatives of the metric in the equations of motion only depend on the metric and the first order partial derivatives of the metric. However,  the equations of motion of the Einstein-Gauss-Bonnet gravity do not have such a property. In general, the equations of motion of Einstein-Gauss-Bonnet gravity are second order full nonlinear equations but not quasi-linear equations. This reason, the Einstein-Gauss-Bonnet gravity and more general Lovelock gravity may have very different behaviors from the Einstein gravity theory in the viewpoint of partial differential equations.

Many physical systems are described by partial differential equations. The Cauchy problem is required to be well-posed by the determinism. The basic causal properties of a system of partial differential equations are determined by its characteristic hypersurfaces~\cite{Vitagliano:2013dra}. As for the Einstein equations, a hypersurface which is characteristic is equivalent to it is null. This means that gravity travels at the speed of light in general relativity. In Einstein-Gauss-Bonnet theory, however, the superluminal propagation of gravitons will arise because of the noncanonical kinetic terms~\cite{Aragone:1987jm}. Along the characteristic hypersurfaces, one can define the effective metric. Izumi has proved some important conclusions by using the method of the characteristics in the Gauss-Bonnet theory~\cite{Izumi:2014loa}. In that paper, He claims that on an evaporating black hole where the geometrical null energy condition is expected not to hold, classical gravitons can escape from the black hole defined with null curves. This kind of study has close relations to the hyperbolicity and causality of the gravitational equations.

In principle, to get the hyperbolicity of the equations of motion for a given gravity theory, we have to study the  leading partial derivative terms of the equations, and get the principle symbol which  is  a matrix in usual. The classification of the partial differential equations depends on the property of this matrix.  However, in gravity theory, usually, people are interested in the propagation of a gravitational fluctuation on a fixed background, i.e, the gravitational perturbation  theory or linearized gravity theory. Although the linearized gravity theory is very different from the full theory,  it really provides some useful information on the hyperbolicity and causality of the full theory~\cite{Reall:2014pwa}. So the study of the gravitational perturbation equation provides a practical way to investigate  the hyperbolicity and causality of the theory.

It is well known that any gravitational perturbation theory is  gauge dependence. The perturbation variables between one gauge and another are related by a gauge transformation. Problems of choosing gauge will be faced when we study the perturbation of a spacetime. One way is to find physically preferred gauges. Another way is to use the gauge-invariant variable, for example, the Kodama-Ishibashi gauge-invariant variables~\cite{Ishibashi:2011ws}. By using these gauge-invariant variables, one can get the master equations which have tensor, vector, and scalar parts~\cite{Ishibashi:2011ws, Dotti:2004sh, Dotti:2005sq, Cai:2013cja}. Based on these, the stability of the higher dimensional black holes such as Gauss-Bonnet black holes are studied by Takahashi and Soda~\cite{Takahashi:2010gz, Takahashi:2009xh, Takahashi:2009dz}. A typical way to calculate the effective metric is using the master equations. This means that one can determine characteristics from the equations of motion of linearized perturbations of a background~\cite{Papallo:2015rna, Reall:2014pwa, Brustein:2017iet}. In the paper~\cite{Papallo:2015rna}, Papallo and Reall work out a speed limit for small black holes and investigate a Shapiro time delay in the Einstein-Gauss-Bonnet theory. In the paper~\cite{Reall:2014pwa}, Reall and Takahashi generalized the result of Izumi~\cite{Izumi:2014loa} to prove that any Killing horizon is a characteristic hypersurface for all gravitational degrees of freedom of a Lovelock theory. They also investigated the hyperbolicity of Ricci flat type N spacetimes and the static, maximally symmetric black hole solutions in the Lovelock theory.  Beside the method by gauge-invariant variables, recently, it has been shown that (weakly coupled) 
Lovelock and Horndeski theories possess a well-posed Cauchy problem base on the 
conventional harmonic gauge or modified harmonic gauge~\cite{Papallo:2017qvl, Kovacs:2020ywu}.

However, unlike the Kodama-Ishibashi formalism in Einstein gravity theory, the gauge-invariant gravitational perturbation in Einstein-Gauss-Bonnet theory and more general Lovelock theory is far from completed --- even the simplest tensor perturbation is still staying at the static spacetimes~\cite{Ishibashi:2011ws}.  So, for black hole spacetimes,  the study of the hyperbolicity and causality of the Einstein-Gauss-Bonnet theory is mainly limited to static cases up to date~\cite{Reall:2014pwa}.  In this paper,  we conquer the general master equation of tensor perturbation for general warped product spacetimes in Einstein-Gauss-Bonnet theory.   Based on this new master equation, we show that effective metric of the tensor perturbation equation defined by Reall in~\cite{Reall:2014pwa} can be generalized to the cases without a static condition.

 Since the new master equation does not depend on the mode expansion (For example, the expansion by  harmonic tensors on a closed manifold with integers $k^2 = \ell(\ell +1)\cdots$ which is proportional to the eigenvalues of Laplacian operator),  the hyperbolicity and causality   can be investigated for any mode of the perturbation.  This 
 generalized the method in~\cite{Reall:2014pwa} in which only large $\ell$ mode has been  studied.  Based on the master equation, we study the hyperbolicity and causality of
all relevant vacuum solutions of the theory. For each solution, we give the exact hyperbolic condition of the tensor perturbation equations. For example, we can give an analytic hyperbolic condition of the tensor perturbation equation on $D=6$ dimensional black holes background.  Our approach can also be applied to dynamical spacetimes, and
Vaidya spacetime have been investigated as an example.

The paper is organized as follows. In section \ref{section2}, we give a brief review on  the Einstein-Gauss-Bonnet theory, and the effective metric defined by Reall, and explain how to get the effective metric from the master equation of tensor part. In section \ref{section3}, we calculate the tensor perturbation equations for a general warped  product spacetime instead of the static solutions. Moreover, we give the condition of keeping the hyperbolicity of the tensor perturbation equations and give the condition of the tensor mode which can travel faster than light. In section \ref{section4}-\ref{section6}, according to the classification of the vacuum solutions in~\cite{Maeda:2007uu}, we check the hyperbolicity of the tensor perturbation equation on three types of spacetimes --- the Boulware-Deser-Wheeler-Cai solution, the Nariai-type spacetime,  and the dimensionally extended constant curvature black hole. We also  check whether there are cases that the graviton travels faster than the light. In the section \ref{section7}, we apply our methods to investigate hyperbolicity of the tensor perturbation equation on Vaidya spacetime which is not a vacuum solution any more. Section \ref{section8} is devoted to conclusion and discussion.

We use the following notion for indices. The capital letters $\left\lbrace M,N,L, P\cdots\right\rbrace$ are the indices for the $D=n+2$-dimensional spacetime. The lowercase letters $a\, ,b$ are the indices for the manifold $(M^2\, ,g_{ab})$, while the lowercase letters $\left\lbrace i\, ,j\, ,k\, ,l\, ,\cdots\right\rbrace$ are the indices for the manifold $(N^n\, ,\gamma_{ij})$. The convention of the  curvature is given by $(\nabla_M\nabla_N-\nabla_N\nabla_M)v_L =R_{MNLP}v^P$, which is the same as in the reference~\cite{Cai:2013cja}. 


\section{Basic theory}
\label{section2}
\subsection{Einstein-Guass-Bonnet theory}
Here, we start by a brief review of Einstein-Gauss-Bonnet gravity with a cosmological constant~\cite{Lovelock:1971yv, Lovelock:1972vz}. The action in the $D$-dimansional spacetime with a metric $g_{MN}$ is given by 
\begin{eqnarray}
	S=\int d^Dx\sqrt{-g}\left[\frac{1}{2\kappa_D^2}\left(R-2\Lambda+\alpha L_{GB}\right)\right]+S_{\text{matter}}\, ,
\end{eqnarray}
where $\kappa_D$ is the coupling constant of gravity, and $R$ and $\Lambda$ are the $D$-dimensional Ricci scalar and the cosmological constant, respectively. $S_{\text{matter}}$ stands for the matter fields. The Gauss-Bonnet term is given by 
\begin{equation}
L_{GB}=R^2-4R_{MN}R^{MN}+R_{MNPQ}R^{MNPQ}\, .
\end{equation}
The symbol $\alpha$ is the coupling constant of the Gauss-Bonnet term. This type of action can be derived from the low-energy limit of heterotic string theory~\cite{Gross:1986iv, Gross:1986mw, Metsaev:1987bc, Metsaev:1987zx}. This reason, $\alpha$ is considered to be positive. The equation of motion of this theory is given by
 \begin{eqnarray}\label{EOM}
G_{MN}+\alpha H_{MN}+\Lambda g_{MN}=\kappa_D^2T_{MN}\, ,
 \end{eqnarray}
where
\begin{eqnarray}
G_{MN}=R_{MN}-\frac{1}{2}g_{MN}R\, ,
\end{eqnarray}
 and
\begin{equation}
\label{GB_tensor}
H_{MN}=2\left[RR_{MN}-2R_{ML}R^{L}_{\ N}-2R^{KL}R_{MKNL}+R_{M}^{\ \ KLP}R_{NKLP}\right]-\frac{1}{2}g_{MN}L_{GB}\, .
\end{equation}
The second order tensor $T_{MN}$,  which can be obtained from $S_{\text{matter}}$, is the energy-momentum tensor for matter fields. In the four-dimensional spacetime, the Gauss-Bonnet term does not make a contribution to the equations of motion since it is identically a total derivative. 
It is worth pointing out that the field equations $(\ref{EOM})$ contain up to the second derivative of the metric just as what the Lovelock theorem said~\cite{Lovelock:1971yv}.

Now, we consider a $D=2+n$-dimensional spacetime $(\mathcal{M}^D\, ,g_{MN})$, which has a local direct product manifold $\mathcal{M}^D\cong M^2\times N^n$ and a metric, 
\begin{equation}
\label{metric}
g_{MN}dx^Mdx^N=g_{ab}(y)dy^a dy^b+r^2(y)\gamma_{ij}(z)dz^idz^j\, ,
\end{equation}
where coordinates $x^M=\{y^1\, ,y^2\, ;z^1\, ,\cdots\, ,z^n\}$. The two element tuple $(M^2\, ,g_{ab})$ forms a two dimensional Lorentzian manifold, and $(N^n,\gamma_{ij})$ is an $n-$dimensional Riemann manifold. 
The metric compatible covariant derivatives associated with $g_{MN}$, $g_{ab}$, and $\gamma_{ij}$ are denoted by $\nabla_M$, $D_a$, and $\hat{D}_i$, respectively. 
In the following discussion, the Riemann manifold $(N^n,\gamma_{ij})$ is assumed to be an Einstein manifold, i.e,
\begin{equation}
\label{Rij1}
\hat{R}_{ij}=(n-1)K\gamma_{ij}\, ,
\end{equation}
where $\hat{R}_{ij}$ is the Ricci tensor of $(N^n,\gamma_{ij})$, and $K=0\, ,\pm 1$. Actually, we will consider more restrictive case in which this Einstein manifold is a maximally symmetric space, and $K$ is the sectional curvature 
of the space.

 An $(n+2)$-dimensional spacetime in GB gravity with the metric $(\ref{metric})$, in which $(N^n,\gamma_{ij})$ is an Einstein manifold, has three groups of  field equations
\begin{eqnarray}\label{EOM1}
\Bigg\{\left[1+2\alpha(n-1)(n-2)\frac{K-(Dr)^2}{r^2}\right]{}^2\!{R}-2(n-1)\frac{\prescript{2}{}{\Box}r}{r}+(n-1)(n-2)\frac{K-(Dr)^2}{r^2}-2\Lambda \nonumber\\
 +4\alpha(n-1)(n-2)
\Bigg[\Big(\frac{\prescript{2}{}{\Box}r}{r}\Big)^2-\frac{(D_aD_br)(D^aD^br)}{r^2}
+\frac{(n-3)(n-4)[K-(Dr)^2]^2}{4r^4}\nonumber\\
-\frac{(n-3)[K-(Dr)^2]}{r^2}\frac{\prescript{2}{}{\Box}r}{r}\Bigg]+\frac{\alpha\hat{C}_{klmn}\hat{C}^{klmn}}{r^4}\Bigg\}\gamma_{ij}-\frac{4\alpha}{r^4}\hat{C}_i{}^{klm}\hat{C}_{jklm}=-2\kappa_{D}^2p\label{EOM_1}\gamma_{ij}\, ,
\end{eqnarray}
\begin{eqnarray}\label{EOM2}
\left[1+2\alpha(n-1)(n-2)\frac{K-(Dr)^2}{r^2}\right]\left(D_aD_br-\frac{1}{2}\prescript{2}{}{\Box}rg_{ab}\right)=-\frac{r}{n}\kappa_{D}^2\left(T_{ab}-\frac{1}{2}T_{cd}g^{cd}g_{ab}\right)\, ,
\label{EOM_2}
\end{eqnarray}
and
\begin{eqnarray}\label{EOM3}
-\frac{\prescript{2}{}{\Box}r}{r}+(n-1)\frac{K-(Dr)^2}{r^2}+\frac{2\alpha(n-1)(n-2)[K-(Dr)^2]}{r^2} \Bigg\{\frac{(n-3)[K-(Dr)^2]}{2r^2}-\frac{\prescript{2}{}{\Box}r}{r}\Bigg\}\nonumber\\
-\frac{2\Lambda}{n}+\frac{\alpha\hat{C}_{ijkl}\hat{C}^{ijkl}}{nr^4}=-\frac{\kappa_{D}^2}{n}g^{ab}T_{ab}\, ,
\label{EOM_3}
\end{eqnarray}
where the energy-momentum tensor $T_{MN}$ has been decomposed into $T_{MN}=\text{diag}\left\lbrace T_{ab}(y),r^2p(y)\gamma_{ij}\right\rbrace\, .$ The Weyl tensor of $(N\, ,\gamma_{ij})$ is denoted by $\hat{C}_{ijkl}$. In the case of vacuum and maximally symmetric $(N\, ,\gamma_{ij})$,  i.e., $T_{MN}=0$ and $\hat{C}_{ijkl}=0$, similar equations can be found in~\cite{Maeda:2007uu}.  It is worth mentioning that the first equation (\ref{EOM1}) comes from the $(ij)-$components of Eq.(\ref{EOM}), the second equation (\ref{EOM2}) comes from the traceless part of the $(ab)-$components of  Eq.(\ref{EOM}), and the third equation (\ref{EOM3}) comes from the trace part of the $(ab)-$components of  Eq.(\ref{EOM}).


\subsection{Effective metrics}

In gravity theory, the so called ``effective metric"  or   ``acoustic metric"  is  important to determine the type of the linearized gravitational equations which are usually second order partial differential equations.  In the following sections, this kind of discussion will be applied to the Einstein-Gauss-Bonnet theory.  By requiring that the effective metric to be Lorentzian,  the equation of the tensor perturbation is hyperbolic in the usual sense.  

 Principally, to study the causality and hyperbolicity of a gravity theory, we have to consider the full nonlinear equations of motion of the theory. However, 
for some background spacetime, the linear perturbation equations can  provide a lot of information on the hyperbolicity or causality of the theory, see~\cite{Reall:2014pwa} for details. 
To understand the effective metric, let us give a brief review on the hyperbolicity of a second order differential equations.

Consider a second order linear differential  equation for an $\mathcal{N}$-dimensional vector $\phi^I$ in a patch of a spacetime with  a coordinate system $\{x^{M}\, ,M=1\, ,\cdots\, ,D\}$:
\begin{eqnarray}
\label{PDE1}
	P^{MNI}{}_J\partial_{M}\partial_N\phi^J +P^{MI}{}_J\partial_M\phi^J+V^{I}{}_J\phi^I=0\, ,
\end{eqnarray}
where $I\, , J=1\, ,\cdots\, ,\mathcal{N}$, and 
$$P^{MN}=(P^{MNI}{}_J)=(P^{NMI}{}_J)=P^{NM}\, ,$$
$$ P^M=(P^{MI}{}_J)\, ,\quad \mathrm{and}\quad V=(V^{I}{}_J)$$ 
are $\mathcal{N}\times \mathcal{N}$ real matrices. 
For a covector $\xi$ at a point $p$, the principal symbol of the equations is given by
\begin{equation}
	P(p,\xi)=P^{MN}(p)\xi_M\xi_N\, .
\end{equation}
This is an $\mathcal{N}\times \mathcal{N}$ matrix and it plays an important role in the classification of the partial differential equations in usual theory of differential equation. The characteristic polynomial $\mathcal{Q}(p,\xi)$ is defined as
\begin{equation}
	\mathcal{Q}(p,\xi)= \det{ P(p,\xi)}\, .
\end{equation}
The hypersurface $\phi=\mathrm{const}$ is a characteristic hypersurface if $\mathcal{Q}(p\, ,d\phi)=0$,  and the normal vector at each point of the characteristic hypersurface is called 
a characteristic direction.  The normal cone at $p$ is defined by the equation $\mathcal{Q}(p\, ,\xi)=0$ at the point $p$, 
and it is important in the discussion of the causality of the theory~\cite{Vitagliano:2013dra, Izumi:2014loa, Reall:2014pwa, Sarbach:2012pr}.

For the gravitational theories with the second order derivatives of metrics, the linearized gravitational equations can be put into the form (\ref{PDE1}). 
Of course, now   $\phi^I$ corresponds to  $h_{MN}=\delta g_{MN}$, and the indices $I$ and $J$ now correspond to $(MN)$ and $(LP)$ respectively.  Actually, this can be realized by rearranging 
$h_{MN}$  into a column vector.   

In general, the characteristic polynomial $\mathcal{Q}(p,\xi)$ is quite complicated.  However, for the  metric (\ref{metric}) with an maximally symmetric space $(N^n\, ,\gamma_{ij})$ (especially for a static case), it is assumed that 
$\mathcal{Q}(p\, ,\xi)$ can be factorized into a form  
\begin{equation}
\label{Qpxi1}
\mathcal{Q}(p\, ,\xi) = \Big(G_S^{M_1N_2}(p)\xi_{M_1}\xi_{N_1}\Big)^{p_s}\cdot \Big(G_V^{M_2N_2}(p)\xi_{M_2}\xi_{N_2}\Big)^{p_V}\cdot \Big(G_T^{M_3N_3}(p)\xi_{M_3}\xi_{N_3}\Big)^{p_T}\, ,
\end{equation}
where $p_S\, ,p_V$, and $p_T$ are the number of degrees of freedom of scalar, vector, and tensor perturbation respectively, and the second order tensor 
$G_S^{MN}$, $G_V^{MN}$, and $G_T^{MN}$ are the so-called
``effective metric" associated with the scalar, vector, and tensor perturbation~\cite{Reall:2014pwa}.  

In the function space formed by  $h_{MN}$,  the scalar, vector, and tensor perturbation of the metric (\ref{metric}) 
belong to three different  subspaces which are orthogonal to each other ~\cite{Ishibashi:2011ws, Ishibashi:2004wx},  so we can 
consider these three kinds of perturbation separately.   In the case only the tensor perturbation is involved, we have
\begin{equation}
\label{Qpxi2}
\mathcal{Q}(p\, ,\xi) = \Big(G_T^{MN}(p)\xi_{M}\xi_{N}\Big)^{p_T}\, .
\end{equation}
Naively speaking, one can say: since the scalar and vector perturbation are shut down, i.e., $p_S=0=p_V$, so Eq.(\ref{Qpxi1}) reduce to the above equation.   According to the paper~\cite{Reall:2014pwa}, a hypersurface is characteristic if, and only if, it is null with respect to the effective metric $G_T^{MN}$.  The tensor perturbation equation is hyperbolic if that the effective metric $G_T^{MN}$ has a Lorentzian signature. 

For several static examples, it has been shown that the factorization (\ref{Qpxi1}) or  (\ref{Qpxi2})  is really  a reasonable assumption~\cite{Reall:2014pwa}.  In this paper, for the tensor perturbation of the general metric (\ref{metric}), we will show that $\mathcal{Q}(p\, ,\xi) $ can be put into the form (\ref{Qpxi2})  even without the static condition if  $(N^n\, ,\gamma_{ij})$ is maximally symmetric. However, without the condition of the maximal symmetry of $(N^n\, ,\gamma_{ij})$, we have no relation (\ref{Qpxi2}) in general.


\section{Hyperbolicity and Causality of General tensor perturbation equations }
\label{section3}

\subsection{General tensor perturbation equations}

Considering a metric perturbation $g_{MN}\to g_{MN}+h_{MN}$, the linear perturbation equations of Gauss-Bonnet gravity are given by 
\begin{equation}
\label{PE0}
	\delta G_{MN}+\Lambda h_{MN}+\alpha \delta H_{MN}=\kappa_{D}^2\delta T_{MN},
\end{equation}
where $\delta G_{MN}$ and $\delta H_{MN}$ are the perturbations of the Einstein tensor and the Gauss-Bonnet tensor of the spacetime $(\mathcal{M}^D,g_{MN})$ respectively, and $\delta T_{MN}$ is the perturbation of the energy-momentum tensor. 

Here, we will get the effective metric for tensor perturbations around a fixed spacetime with the metric (\ref{metric}). The tensor perturbations around backgrounds (with the Einstein manifold $(N\, ,\gamma_{ij})$) are transverse and traceless part of $h_{ij}$~\cite{Ishibashi:2011ws}, i.e. $h_{ij}=h_{ij}^{\mathrm{TT}}$, which means that we are considering perturbation $h_{MN}$ which satisfies
\begin{eqnarray}
\label{hab1}
	h_{ab}=0\, ,\quad h_{ai}=0\, ,\quad h=h^i_{\ i}=0\, ,\quad \hat{D}_ih^{ij}=0\, .
\end{eqnarray}
It is well known that the above tensor perturbation  of the spacetime with the metric (\ref{metric}) is  gauge invariant.  Of course the  scalar and vector perturbations are quite different, to get the gauge invariant perturbation variables, one has to consider the combinations of different parts of $h_{MN}$.

Substituting $h_{MN}$ (satisfying (\ref{hab1})) into Eq.(\ref{PE0}), 
after lengthy and tedious  calculation, the tensor perturbation equation has the following form 
\begin{equation}
\label{PMNIJ1}
\Big(P^{ab}{}_{ij}{}^{kl} D_aD_b + P^{mn}{}_{ij}{}^{kl} \hat{D}_m\hat{D}_n + P^{a}{}_{ij}{}^{kl} D_a + V_{ij}{}^{kl}\Big)\Big(\frac{h_{kl}}{r^2}\Big) = -\frac{2\kappa_D^2}{r^2}\delta T_{ij}\, ,
\end{equation}
where the detailed expressions for the coefficients $P^{ab}{}_{ij}{}^{kl}$, $P^{mn}{}_{ij}{}^{kl}$, $P^{a}{}_{ij}{}^{kl}$, $ V_{ij}{}^{kl}$ can be found in  Appendix.A. When $(N\, ,\gamma_{ij})$ is maximally symmetric, the Weyl tensor $\hat{C}_{ijkl}$ is vanishing, and  we have
\begin{eqnarray}
&& P^{ab}{}_{ij}{}^{kl} = P^{ab}\delta_i{}^k \delta_{j}{}^l\, ,\nonumber\\
&&P^{mn}{}_{ij}{}^{kl} = P^{mn}\delta_i{}^k \delta_{j}{}^l\, ,\nonumber\\
&&P^{a}{}_{ij}{}^{kl} = P^a \delta_i{}^k \delta_{j}{}^l\, ,\nonumber\\
&& V_{ij}{}^{kl} = V\delta_i{}^k \delta_{j}{}^l\, ,
\end{eqnarray}
where $P^{ab}$, $P^{mn}$, $P^a$, $V$ have following forms
\begin{eqnarray}
\label{Pab}
P^{ab}= g^{ab} + 4\alpha (n-2) \left\lbrace \frac{D^aD^br}{r}+\left[\frac{1}{2}(n-3)\frac{K-(Dr)^2}{r^2}-\frac{\prescript{2}{}{\Box}r}{r}\right]g^{ab}\right\rbrace\, ,
\end{eqnarray}
\begin{eqnarray}
\label{PijQ1}
P^{mn}=\Bigg\{1
+2 \alpha\left[{}^2\!{R}- \frac{2(n-3)\prescript{2}{}{\Box}r}{r}+ (n-3)(n-4)\frac{K-(Dr)^2}{r^2}\right] \Bigg\}\frac{\gamma^{mn}}{r^2}\, ,
\end{eqnarray}
\begin{eqnarray}
\label{Pa123}
P^{a}&=&n\frac{D^ar}{r} +  2(n-2) \alpha \Bigg\{4\frac{D^aD^br}{r}+\Big[{}^2\!{R}
-2(n-1)\frac{\prescript{2}{}{\Box}r}{r}\nonumber\\ 
&&+(n-2) (n-3)\frac{K-(Dr)^2}{r^2}\Big]g^{ab}\Bigg\}\frac{D_br}{r}\, ,
\end{eqnarray}
and
\begin{eqnarray}
\label{V123}
V&=&\prescript{2}{}{R}-2(n-1)\frac{\prescript{2}{}{\Box}r}{r}+\frac{n(n-3)K}{r^2}-\frac{(n-1)(n-2)(Dr)^2}{r^2}-\Lambda\nonumber\\
&+&\alpha\Bigg\{-4(n-1)(n-2)\frac{(D^aD^br)(D_aD_br)}{r^2}+4(n-1)(n-2)\left(\frac{\prescript{2}{}{\Box}r}{r}\right)^2\nonumber\\
&+&2n(n-3)\frac{K\cdot {}^2\!R}{r^2}-2(n-1)(n-2)\frac{(Dr)^2\cdot {}^2\!{R}}{r^2}-4n(n-3)^2\frac{K\cdot\prescript{2}{}{\Box}r}{r^3}\nonumber\\
&+&4(n-1)(n-2)(n-3)\frac{(Dr)^2\cdot\prescript{2}{}{\Box}r}{r^3}-2n(n-3)^2(n-4)\frac{K\cdot(Dr)^2}{r^4}\nonumber\\
&+&(n-3)(n-4)(n^2-3n-2)\frac{K^2}{r^4}+(n-1)(n-2)(n-3)(n-4)\left[\frac{(Dr)^2}{r^2}\right]^2\Bigg{\}}\, .
\end{eqnarray}
So, when $(N\, ,\gamma_{ij})$ is maximally symmetric,  the tensor perturbation equation can written as
\begin{eqnarray}
\label{PE1}
\Big(P^{ab} D_aD_b + P^{kl} \hat{D}_k\hat{D}_l  + P^a D_a  + V \Big)\Big(\frac{h_{ij}}{r^2}\Big) = -\frac{2\kappa_{D}^2}{r^2} \delta T_{ij}\, .
\end{eqnarray}
Obviously, this equation reduces to the tensor perturbation equation in Einstein gravity theory when $\alpha$ is vanishing~\cite{Ishibashi:2011ws, Cai:2013cja}. In the static case, one can check the above equations exactly reduce
to the one in reference~\cite{Takahashi:2010ye}.

\subsection{Hyperbolicity of the tensor perturbation equations}

Comparing Eq.(\ref{PMNIJ1}) and Eq.(\ref{PDE1}), it is easy to find that indices $I$ and $J$ correspond to $(ij)$ and $(kl)$. We can rearrange $h_{ij}=h_{ij}^{\mathrm{TT}}$ into a column vector $\phi^I$ such that
$\delta_i{}^k\delta_j{}^l$ in Eq.(\ref{PMNIJ1}) can be expressed as $\delta_I{}^J$.  So the characteristic polynomial $\mathcal{Q}(p,\xi)$ now has a form (\ref{Qpxi2}), i.e.,
$$\mathcal{Q}(p,\xi) =  \Big(P^{MN}(p)\xi_{M}\xi_{N}\Big)^{p_T}\, ,$$
where $P^{MN}$ can be put into a matrix form
\begin{equation}
\label{PMN5}
(P^{MN})= \begin{bmatrix}
		P^{ab} & 0\\
		0 & \displaystyle \frac{Q}{r^2}\gamma^{ij}
	\end{bmatrix}\, .
\end{equation}
Here $P^{ab}$ has been defined in Eq.(\ref{Pab}), and $Q$ is the coefficient of $\gamma^{ij}/r^2$ in Eq.(\ref{PijQ1}), i.e., 
\begin{equation}
\label{Q}
Q=1+2 \alpha\left[\prescript{2}{}{R}- \frac{2(n-3)\prescript{2}{}{\Box}r}{r}+ (n-3)(n-4)\frac{K-(Dr)^2}{r^2}\right] \, .
\end{equation}
Since $h_{ij}=h_{ij}^{\mathrm{TT}}$, the dimension of the vector $\phi^I$ is the number of degrees of freedom of gravitational perturbation, i.e., we have
$$p_T = \frac{1}{2}D(D-1) -D\, .$$
Thus the effective metric  is nothing but $P^{MN}$ in Eq.(\ref{PMN5}), i.e., we have $$G_T^{MN}=P^{MN}\, .$$  It should be noted here: the effective metric is not so simple 
if that the Einstein manifold $(N\, ,\gamma_{ij})$ is not maximally symmetric, see Appendix A. From now on, we only consider the cases with maximal symmetry.

The effective metric $P^{MN}$ should be Lorentzian to  maintain the hyperbolicity of the theory.  Therefore, we have to set 
\begin{equation}
\label{PQ11}
P=\det(P^{ab})<0\, ,\quad \mathrm{and} \quad Q>0\, .
\end{equation}
By using the equations of motion and the  null frame $\{\ell^a\, ,n^a\}$, we have (see Appendix.B) 
\begin{eqnarray}
\label{P_nomal}
P=\det(P^{ab})=-\Bigg\{1-2\alpha(n-2)\left[\frac{\prescript{2}{}{\Box}r}{r}-(n-3)\frac{K-(Dr)^2}{r^2}\right]\Bigg\}^2 \nonumber\\
+16\alpha^2(n-2)^2\kappa_{D}^4  (T_{nn}T_{\ell\ell} )  \Bigg\{n \left[1+2\alpha(n-1)(n-2)\frac{K-(Dr)^2}{r^2}\right]\Bigg\}^{-2} \, ,
\end{eqnarray}
where   $T_{\ell\ell}=T_{ab}\ell^a\ell^b$ and $T_{nn}=T_{ab}n^an^b$ are the components for the energy-momentum tensor along the null directions.  Of course, the above equation is valid only in the case where 
$$1+2\alpha(n-1)(n-2)\frac{K-(Dr)^2}{r^2}\ne 0\, .$$ 
It will be discussed separately when the above condition is not satisfied (see section.\uppercase\expandafter{\romannumeral6}).

Eq.(\ref{P_nomal}) suggests that $P$ is always negative or vanishing  for vacuum solutions.  When matter fields are present and satisfy null energy condition, $P$ might be positive. 
However, for the radiation matter with a null direction $\ell^a$ or $n^a$, $P$ is also nonpositive. This happens in Vaidya spacetimes.

\subsection{Causality of the tensor perturbation equations}
Once the system is hyperbolic, we can discuss the  speed of a  gravitational fluctuation.  In other words, we can  check whether gravity travels faster than light or not.  For simplicity, we suppose 
\begin{eqnarray}\label{static}
	D^aD^br-\frac{1}{2}\prescript{2}{}{\Box}rg^{ab}=0\, .
\end{eqnarray}
This condition implies that $K^a=\epsilon^{ab}D_b r$ corresponds to a Killing vector field of the spacetime (where $\epsilon_{ab}$ is the volume element of $(M^2, g_{ab})$).  
If gravity travels faster than light, we have
\begin{eqnarray}
\label{condition}
	g^{MN}\xi_M\xi_N<0\, ,
\end{eqnarray}
where $\xi_M$ satisfies the following condition
\begin{eqnarray}
\label{PMN123}
	P^{MN}\xi_M\xi_N=0\, .
\end{eqnarray}
This means that $\xi^M$ is the a characteristic direction, and normal cone by $g^{MN}$ is smaller than the normal cone by $P^{MN}$~\cite{Reall:2014pwa}.
Substituting the $P^{MN}$ in Eq.(\ref{PMN5}) into Eq.(\ref{PMN123}) , we have
\begin{eqnarray}
\label{condition2}
	\left\lbrace1-2\alpha(n-2)\left[\frac{\prescript{2}{}{\Box}r}{r}-(n-3)\frac{K-(Dr)^2}{r^2}\right]\right\rbrace g^{ab}\xi_a\xi_b+Qg^{ij}\xi_i\xi_j=0\, .
\end{eqnarray}
Since  $g^{ij}=\gamma^{ij}/r^2$ is positive defined, Eq.(\ref{condition}) implies that
\begin{eqnarray}
	g^{ab}\xi_a\xi_b<0\, .
\end{eqnarray}
Only in the case $Q>0$, it makes sense of discussing the velocity of the graviton.   So we have
\begin{eqnarray}
	1-2\alpha(n-2)\left[\frac{\prescript{2}{}{\Box}r}{r}-(n-3)\frac{K-(Dr)^2}{r^2}\right]>0\, .
\end{eqnarray}
For convenience, we define
\begin{eqnarray}
	I=1-2\alpha(n-2)\left[\frac{\prescript{2}{}{\Box}r}{r}-(n-3)\frac{K-(Dr)^2}{r^2}\right].
\end{eqnarray}
Therefore, from Eqs.($\ref{condition}$) and (\ref{condition2}), we arrive at
\begin{eqnarray}
	Q-I>0\, .
\end{eqnarray}
From the expressions of $Q$ and $I$, it is not hard to find that $Q-I$ is proportional to $\alpha$, so we can define 
$$Q-I=2\alpha J\, ,$$ 
where $J$ is given by
\begin{eqnarray}
	J={}^2\!{R}-(n-4)\frac{\prescript{2}{}{\Box}r}{r}-2(n-3)\frac{K-(Dr)^2}{r^2}\, .
\end{eqnarray}
It is easy to find that $I$ approaches $Q$ if we take the limit $\alpha\rightarrow 0$. So, under this limit,  the gravitational wave travels at  a speed of light.
In conclusion, if gravity travels faster than light, we have the following three conditions
\begin{equation}
Q=1+2 \alpha\left[{}^2\!{R}- \frac{2(n-3)\prescript{2}{}{\Box}r}{r}+ (n-3)(n-4)\frac{K-(Dr)^2}{r^2}\right]>0
\label{IEQ1}\, ,
\end{equation}
\begin{equation}
I=1-2\alpha(n-2)\left[\frac{\prescript{2}{}{\Box}r}{r}-(n-3)\frac{K-(Dr)^2}{r^2}\right]>0
\label{IEQ2}\, ,
\end{equation}
and
\begin{equation}
J= {}^2\!{R}-(n-4)\frac{\prescript{2}{}{\Box}r}{r}-2(n-3)\frac{K-(Dr)^2}{r^2}> 0
 \label{IEQ3}\, .
\end{equation}
In the following sections, we will discuss the hyperbolicity and causality of the tensor perturbation equations for exact solutions in Einstein-Gauss-Bonnet gravity theory.
We will investigate the above three inequalities on these exact spacetimes.


\section{Boulware-Deser-Wheeler-Cai solution}
\label{section4}
The static vacuum solution in Einstein-Gauss-Bonnet theory has been found long time ago by Boulware and Deser~\cite{Boulware:1985wk}  and Wheeler~\cite{Wheeler:1985nh}. The solutions have been extended to the case with a cosmological constant  by Cai~\cite{Cai:2001dz} about twenty years ago. The metric of the solution is given by
\begin{equation}
\label{solution1}
ds^2=-F(r)dt^2+F^{-1}(r)dr^2+r^2\gamma_{ij}dz^idz^j\, ,
\end{equation}
where
\begin{equation}
F(r)=K+\frac{r^2}{2\tilde{\alpha}}\left[1\mp\sqrt{1+4 \tilde{\alpha} \Bigg(\frac{2 \kappa_{D}^2 M}{nV_{n}^K}\frac{1}{r^{n+1 }}+\tilde{\Lambda}}\Bigg)\right]\, .
\end{equation}
In the abvoe equation, $``-"$ corresponds to the so-called ``General Relativity branch" which reduces to the solution in general relativity when $\alpha$ approaches to zero, while $``+"$ corresponds to ``Gauss-Bonnet branch" which has no general relativity limit as $\alpha$ approaches to zero.
The parameter $M$ is mass parameter, and $\tilde{\alpha}=(n-1)(n-2)\alpha$ and $\tilde{\Lambda}=2\Lambda/(n(n+1))$.
In the following discussion, It is convenient to define $\tilde{M}$ as
$$\tilde{M}=\frac{2 \kappa_{D}^2M}{nV_{n}^K}\, ,$$
where $V_n^K$ is the volume of the maximally symmetric space $(N\, ,\gamma_{ij})$ with a unit radius. 
Since it is a vacuum solution, from Eq.(\ref{P_nomal}), we always have $P<0$ (excluding some special values of $P=0$ which form a zero measure set). Therefore, the hyperbolicity is determined by the sign of $Q$. Some calculation shows
\begin{equation}
Q=1-2 \alpha\left[F^{\prime\prime}(r)+\frac{2(n-3)F^{\prime}(r)}{r}- (n-3)(n-4)\frac{K-F(r)}{r^2}\right]\, . 
\end{equation}
This result is the same as the one  in  reference~\cite{Brustein:2017iet} when $n=3$ and $n=4$. 

Another important thing is to study the existence of the superluminal modes. Before the detailed discussion, we give three useful formulas which can be expressed as
\begin{equation}
Q_{\text{GR}}+Q_{\text{GB}}=0\, ,
\end{equation}
\begin{equation}
I_{\text{GR}}+I_{\text{GB}}=0\, ,
\end{equation}
\begin{equation}
J_{\text{GR}}+J_{\text{GB}}=0\, ,
\end{equation}
where GR stands for the General Relativity branch while GB stands for the Gauss-Bonnet branch. It is easy to get the above equations, and we will not give the proof here.  These results imply that the discussion on the GR branch is enough.
So, in this paper, we only consider the solution which can reduce to the one in general relativity under the limit of $\alpha\to 0$.

The details of the solution (\ref{solution1}) is complicated. For different values of the parameters, the solution might be a globally regular solution,  a black hole,  a naked singularity, or a branch singularity. The classification of this solution
has been done in~\cite{Torii:2005xu} and references therein.

\subsection{Black hole solutions}

In this subsections, we show  that the hyperbolicity is broken outside the event horizon in some dimensions, for example  $D=6$, and this is consistent with the results in~\cite{Reall:2014pwa}.
Beside this, we give the precise conditions when  the hyperbolicity is broken outside the event horizon. 

For simplicity, when the event horizon is present, we introduce three dimensionless quantities, $x$, $a$, and $\lambda$ 
\begin{equation}
\label{xalambda}
x=\frac{r_+}{r}\, , \qquad a=\frac{\tilde{\alpha}}{r_+^2}\, ,\qquad \lambda=r_+^2\tilde{\Lambda}\, ,
\end{equation} 
where $r_+$ is the radius of the outermost event horizon. Here, we have assumed $\alpha>0$, so $a$ is always positive. By these definitions,  $Q$, $I$, and $J$ can be expressed  as
\begin{equation}
Q=1-\frac{2ax^2}{(n-1)(n-2)}\Big[x^2F_{xx}-2(n-4)xF_x-(n-3)(n-4)(K-F(x))\Big]\, ,
\end{equation}
\begin{equation}
	I=1+\frac{2ax^2}{(n-1)}\left[xF_x+(n-3)(K-F(x))\right]\, ,
\end{equation}
\begin{equation}
	J=\frac{1}{r^2}\left[-x^2F_{xx}+(n-6)xF_x-2(n-3)(K-F(x))\right]\, ,
\end{equation}
where 
\begin{equation}
\label{Fx1}
F(x)=K+\frac{1}{2ax^2}\Big\{1-\sqrt{1+4a\lambda+\left[(1+2Ka)^2-1-4a\lambda\right]x^{n+1}}\Big\}\, .
\end{equation}
To simplify the discussion, it is convenient to define
\begin{equation}
\hat{J} = r^2 J = -x^2F_{xx}+(n-6)xF_x-2(n-3)(K-F(x))\, .
\end{equation}
Note that $J$ and $\hat{J}$ have the same sign. Here the $F_x$ denotes the derivative of $F(x)$ with respect to $x$.  It should be noted here that $F_x$ is negative at $x=1$ because  $r_+$ is the radius of the outermost event horizon of the black hole.

\subsubsection{$\Lambda=0$ and $K=1$ case}

For $n=3$, when $\tilde{M}>\tilde{\alpha}$, there is a black hole horizon. For $n\ge4$, when $\tilde{M}>0$, there is a black hole horizon~\cite{Torii:2005xu}.  
Here, we will investigate the sign of $Q$ outside the (outmost) event horizons of the black holes. In these regions of the spacetimes, $x\in\left[0,1\right]$, and $Q$
can be written as
\begin{equation}
\label{Q12}
Q=\frac{(n-3)(n-5)y^4+2(n+1)(2n-3)y^2-(n+1)^2}{4(n-1)(n-2)y^3}\, ,
\end{equation}
where 
$$y=[1+4a(1+a)x^{n+1}]^{\frac{1}{2}}\, .$$
Obviously, we have  $y\in\left[1,1+2a\right]$.  Since the denominator of $Q$ in Eq.(\ref{Q12}) is always positive, we will focus on  the numerator of $Q$, denoted by $\hat{Q}$,    given by
\begin{equation}
\label{hatQ1234}
	\hat{Q}(y)=(n-3)(n-5)y^4+2(n+1)(2n-3)y^2-(n+1)^2\, .
\end{equation}
We investigate the sign of $\hat{Q}$ in the following discussion. The $n=3\, ,4\, ,5$ and $n\ge 6$ will be discussed separately. 

For $n=3$, it is easy to find that
\begin{equation}
\hat{Q}(y)=24y^2-16\ge24-16=8>0\nonumber\, .
\end{equation}
So in $D=5$, the tensor perturbation is hyperbolic  outside the event horizon of the spacetime.

For $n=4$, we have
\begin{equation}
\hat{Q}(y)=-y^4+50y^2-25\nonumber\, .
\end{equation}
This equation implies
\[\begin{cases}
\hat{Q}> 0\, ,& y\in\left[1,\sqrt{25+10\sqrt{6}}\right)\, ,\\
\hat{Q}<0\, , & y\in \left(\sqrt{25+10\sqrt{6}},+\infty\right)\, .
\end{cases}\]
Therefore, $\hat{Q}$  is positive for all $y \in[1,1+2a]$ when $$1+2a<\sqrt{25+10\sqrt{6}}\, .$$ However,  $\hat{Q}$  is negative if $y\in (\sqrt{25+10\sqrt{6}},1+2a] $ when $$1+2a>\sqrt{25+10\sqrt{6}}\, ,$$ and $\hat{Q}$ is positive if $y\in [1,\sqrt{25+10\sqrt{6}})$.

For $n=5$,  we find
\begin{equation}
\hat{Q}=84y^2-36\ge84-36=48>0\nonumber\, .
\end{equation} 
So, in the case of $D=7$, the equation is always hyperbolic outside the horizon.

For $n\ge6$,  from the expression (\ref{hatQ1234}), it is not hard to find that
$\hat{Q}(y)$ has only one zero point $t$ in $\left(0,+\infty\right)$, where 
\begin{equation}
\label{t1}
	t=\sqrt{\frac{(n+1)\left[\sqrt{5n^2-20n+24}-(2n-3)\right]}{(n-3)(n-5)}}\, .
\end{equation} 
However, it is not hard to prove that $t<1$. This means that  $Q$ is positive for all $y\in [1,1+2a ]$. 

The above discussion shows that tensor perturbation equation is hyperbolic outside the event horizon for all possible parameter $a$ if $D\ne 6$.  The case of $D=6$ or $n=4$ is very special, the hyperbolicity is maintained outside the event horizon only when 
$$a=\frac{\tilde{\alpha}}{r_+^2} < \frac{1}{2}\Big(\sqrt{25 + 10 \sqrt{6}} -1\Big)\approx 3.0176.$$
This suggests that the radius of  the event horizon of the  black hole  has to satisfy 
$$r_+> \sqrt{\tilde{\alpha}/3.0176}\, .$$ So, to ensure the hyperbolicity of the equation,  the black hole can not be too small.

\subsubsection{$\Lambda<0$ and $K=1$ case}

For $n=3$, when $\tilde{M}> \tilde{\alpha}$, there is a black hole horizon. For $n\ge4$, when $\tilde{M}>0$, there is a black hole horizon~\cite{Torii:2005xu}.
Outside the event horizon, we have $x\in\left[0,1\right]$, and $Q$ has a form
\begin{eqnarray}
\label{Q13}
Q(y)= \frac{(n-3)(n-5)y^4+2(n+1)(2n-3)(1+4a\lambda)y^2-(n+1)^2(1+4a\lambda)^2}{4(n-1)(n-2)y^3}\, ,
\end{eqnarray}
where
\begin{equation}
y=[1+4a\lambda+4a(-\lambda+1+a)x^{n+1}]^{\frac{1}{2}}\, .
\end{equation}
Since $x\in [0,1 ]$, we have  $y\in [\sqrt{1+4a\lambda},1+2a]$.

\begin{itemize}

\item[(i).] Firstly, let us consider the case with $1+4a\lambda>0$.   Similar to the case with $\Lambda=0$,  the numerator of $Q$ in Eq.(\ref{Q13})  can be defined as 
\begin{equation}
\label{hatQ13}
\hat{Q}(y)=(n-3)(n-5)y^4+2(n+1)(2n-3)(1+4a\lambda)y^2-(n+1)^2(1+4a\lambda)^2\, .
\end{equation}
For $n=3$, we have
\begin{eqnarray}
		\hat{Q}(y)=24(1+4a\lambda)y^2-16(1+4a\lambda)^2\ge 8(1+4a\lambda)^2>0\nonumber\, .
\end{eqnarray}
For $n=4$, we have
\begin{eqnarray}
\label{Q41}
	\hat{Q}(y)=-y^4+50(1+4a\lambda)y^2-25(1+4a\lambda)^2\nonumber\, .
\end{eqnarray}
From this equation, we get
\[\begin{cases}
\hat{Q}> 0\, ,& y\in\left[\sqrt{1+4a\lambda}\, ,\sqrt{1+4a\lambda}\cdot\sqrt{25+10\sqrt{6}}\right)\, ,\\
\hat{Q}<0\, , & y\in \left(\sqrt{1+4a\lambda}\cdot\sqrt{25+10\sqrt{6}}\, ,+\infty\right)\, .
\end{cases}\]
Thus, we have $\hat{Q}>0$ for all  $y\in [\sqrt{1+4a\lambda}\, ,1+2a]$ when 
\begin{equation}
\label{n4condition1}
1+2a<\sqrt{1+4a\lambda}\cdot\sqrt{25+10\sqrt{6}}\, .
\end{equation}
However, we have $\hat{Q}<0$ if $y\in (\sqrt{1+4a\lambda}\cdot\sqrt{25+10\sqrt{6}},1+2a]$ when
$$1+2a>\sqrt{1+4a\lambda}\cdot\sqrt{25+10\sqrt{6}}\, ,$$
and $\hat{Q}>0$ if $y\in [\sqrt{1+4a\lambda}\, , \sqrt{1+4a\lambda} \cdot \sqrt{25+10\sqrt{6}})$.

For $n=5$, obviously we have 
\begin{eqnarray}
\hat{Q}(y)=84(1+4a\lambda)y^2-36(1+4a\lambda)^2\ge48(1+4a\lambda)^2>0\nonumber\, .
\end{eqnarray}

For $n\ge6$, from the expression of $\hat{Q}(y)$ in (\ref{hatQ13}),  we find that
$\hat{Q}(y)$ has only one  zero ponit $(\sqrt{1+4a\lambda})~t$ in $\left(0,+\infty\right)$, where $t$ is given by Eq.(\ref{t1}).
Noted that $t<1$, we have $(\sqrt{1+4a\lambda} ) t<\sqrt{1+4a\lambda}$. So, we always have $Q>0$ when $n\ge 6$.

So, as the case $\Lambda<0$ and $K=1$, the $D=6$ or $n=4$ is special. If we hope the hyperbolicity is present outside the event horizon of the black hole,
we have to impose a condition on $a$ and $\lambda$, i.e., the inequality (\ref{n4condition1}), see  Fig.\ref{fig1}.  The hyperbolicity is broken above the  solid black line.
\begin{figure}[htbp]
\centering
\includegraphics[width=3.0 in]{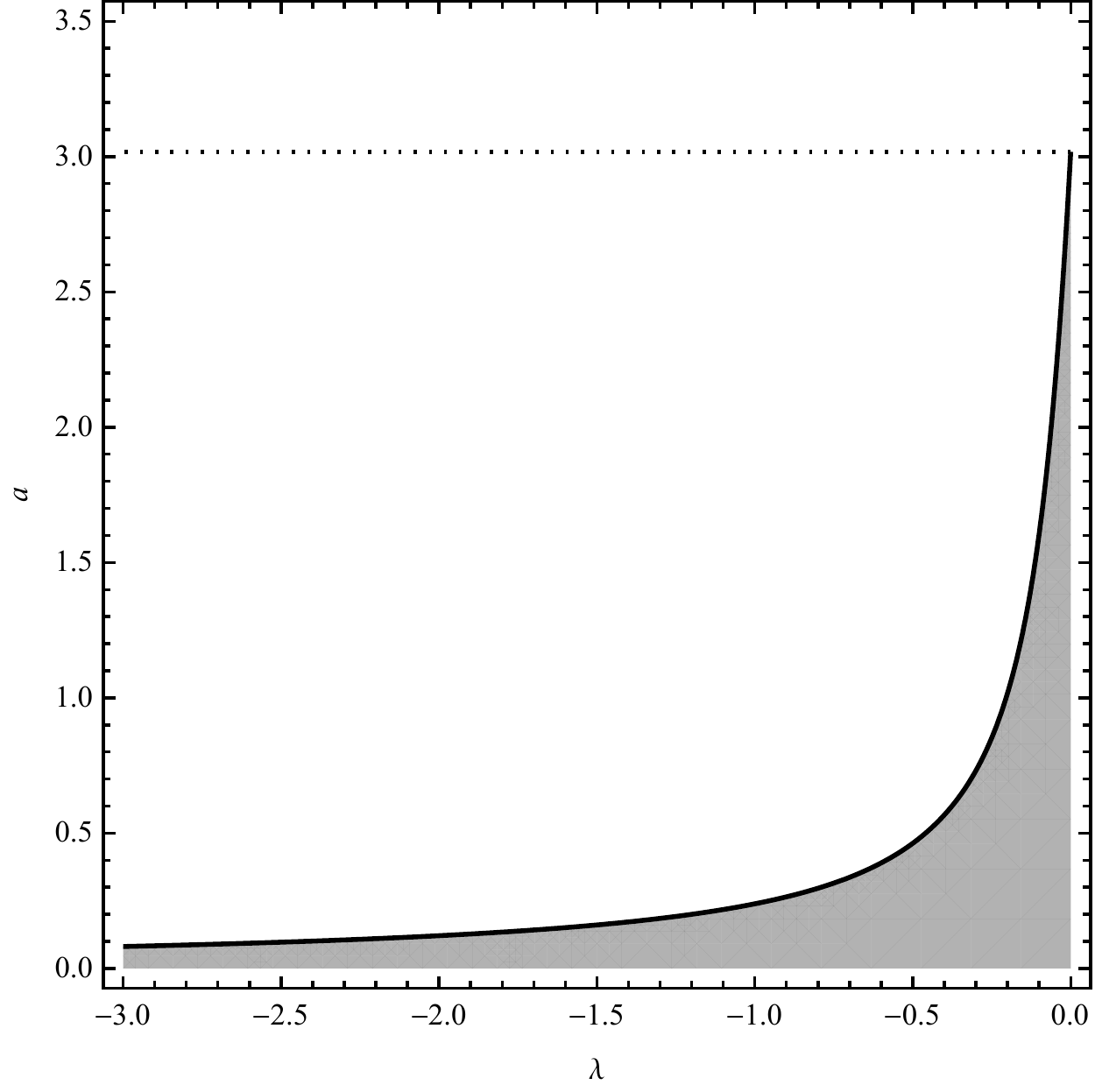}
\caption{$n=4$, the case for $\Lambda<0$ and $K=1$. In the shadow region, inequalities  (\ref{n4condition1})  and $1+4a\lambda>0$ are satisfied. 
For a given $\lambda$, we have an upper bound for $a$. The dotted line is for $a=3.0176$, and this corresponds to the upper bound of $a$ in the case with $\Lambda=0$ and $K=1$.    }
\label{fig1}
\end{figure}

\item[(ii).] Secondly, when $1+4a\lambda=0$, from $x\in\left(0,1\right]$ we have $y\in\left(0,1+2a\right]$, and $Q$ has a simple form
\begin{equation}
\label{Q5}
Q=\frac{(n-3)(n-5)y}{4(n-1)(n-2)}\, .
\end{equation}
It is obvious that $Q=0$ when $n=3\, ,5$, $Q<0$ when $n=4$, and $Q>0$ when $n\ge6$.

\item[(iii).] Finally,  in the case with $1+4a\lambda<0$,  it is not hard to find 
$$x\in\left(\left[\frac{-(1+4a\lambda)}{4a(-\lambda+1+a)}\right]^{\frac{1}{n+1}},1\right]\, .$$
 then $y\in\left(0,1+2a\right]$. Actually, physically, this case is not so interesting because  the solution has some (branch)  singularity when $y$ approaches to $0$. However, 
 it is also curious to us to that the possibility of the existence of some wave equation on this strange background. From the expression of $\hat{Q}(y)$ in Eq.(\ref{hatQ13}) and 
 the condition $1+4a\lambda<0$, it is easy to find we have $\hat{Q}(y) <0$ for $n=3\, ,4\, ,5$. So $Q(y)<0$ when $n\le 5$. The linearized gravitational wave does not exist in 
when $D\le 7$.

For $n\ge6$, $\hat{Q}(y)$ has only one zero point $s$ in $\left(0,+\infty\right)$, where 
\begin{equation}
\label{s1}
s=\sqrt{-(1+4a\lambda)\frac{(n+1)\left[\sqrt{5n^2-20n+24}+(2n-3)\right]}{(n-3)(n-5)}}\, .
\end{equation}
This implies we always have $Q<0$ in $\left(0,1+2a\right]$ when $1+2a<s$. However, we can get  $Q>0$ in $\left(s,1+2a\right]$  when $1+2a>s$, and $Q$ is still negative in $(0,s)$. So, for this background with a branch singularity,
we can get some wave equations when $D\ge 8$.
\end{itemize}

\subsubsection{$\Lambda<0$ and $K=0$ case}

For any $n\ge3$, when $\tilde{M}>0$, there is a black hole horizon~\cite{Torii:2005xu}. Consider the region outside the event horizon, then we have $x\in\left[0,1\right]$.  
By defining
$$y=\sqrt{1+4a\lambda-4a\lambda x^{n+1}}\, ,$$
the $Q$ and $\hat{Q}$ have the same forms
as in Eq.(\ref{Q13}) and (\ref{hatQ13}).

\begin{itemize}
\item[(i).] In the case with $1+4a\lambda>0$, we have $y\in\left[\sqrt{1+4a\lambda},1\right]$. Similar to the case with $K=1$,  $n=3\, ,4\, , 5$, and $n\ge 6$ will be discussed separately. 

For $n=3$, we have
\begin{equation}
	\hat{Q}(y)=24(1+4a\lambda)y^2-16(1+4a\lambda)^2\ge8(1+4a\lambda)^2>0\nonumber\, .
\end{equation}
For $n=4$, we have
\begin{eqnarray}
	\hat{Q}(y)=-y^4+50(1+4a\lambda)y^2-25(1+4a\lambda)^2\nonumber\, .
\end{eqnarray}
This equation implies
\[\begin{cases}
\hat{Q}> 0\, ,& y\in\left[\sqrt{1+4a\lambda}\, ,\sqrt{1+4a\lambda}\cdot\sqrt{25+10\sqrt{6}}\right)\, ,\\
\hat{Q}<0\, , & y\in \left(\sqrt{1+4a\lambda}\cdot\sqrt{25+10\sqrt{6}}\, ,+\infty\right)\, .
\end{cases}\]
So we have $\hat{Q}>0$ for all $y$ in  $[\sqrt{1+4a\lambda}\, ,1]$ if 
\begin{equation}
\label{n4condition2}
1<\sqrt{1+4a\lambda}\cdot\sqrt{25+10\sqrt{6}}\, .
\end{equation}
However, $\hat{Q}<0$ for $y$ in $(\sqrt{1+4a\lambda}\cdot\sqrt{25+10\sqrt{6}},1]$ if
$$1>\sqrt{1+4a\lambda}\cdot\sqrt{25+10\sqrt{6}}\, .$$ 
Of course, now we also have $\hat{Q}>0$ for $y$ in $[\sqrt{1+4a\lambda}\, ,\sqrt{1+4a\lambda}\cdot \sqrt{25+10\sqrt{6}})$.

For $n=5$, we have 
\begin{eqnarray}
	\hat{Q}(y)=84(1+4a\lambda)y^2-36(1+4a\lambda)^2\ge48(1+4a\lambda)^2>0\nonumber\, .
\end{eqnarray}

For $n\ge6$, from the expression (\ref{hatQ13}), we find that
$\hat{Q}(y)$ has only one  zero ponit $(\sqrt{1+4a\lambda})t$ in $\left(0,+\infty\right)$, where $t$ is the same as the one in Eq.(\ref{t1}). Noted  that $t<1$, we have $(\sqrt{1+4a\lambda})t<\sqrt{1+4a\lambda}$. So we have $Q>0$.

Above discussions show that $D=6$ is special, one might get negative $Q$ if the condition (\ref{n4condition2}) is broken. The details can be found in Fig.\ref{fig2}.  Obviously,  for a given $\lambda$, we have an upper bound of $a$. However,
this upper bound approaches infinity when $\lambda$ approaches zero.   This is very different from the case with $K=1$. Physically, this suggests: by choosing the value of $\lambda$,  the black hole  can be arbitrary small without breaking the hyperbolicity of the tensor perturbation equation. 

\begin{figure}[htbp]
\centering
\includegraphics[width=3.0 in]{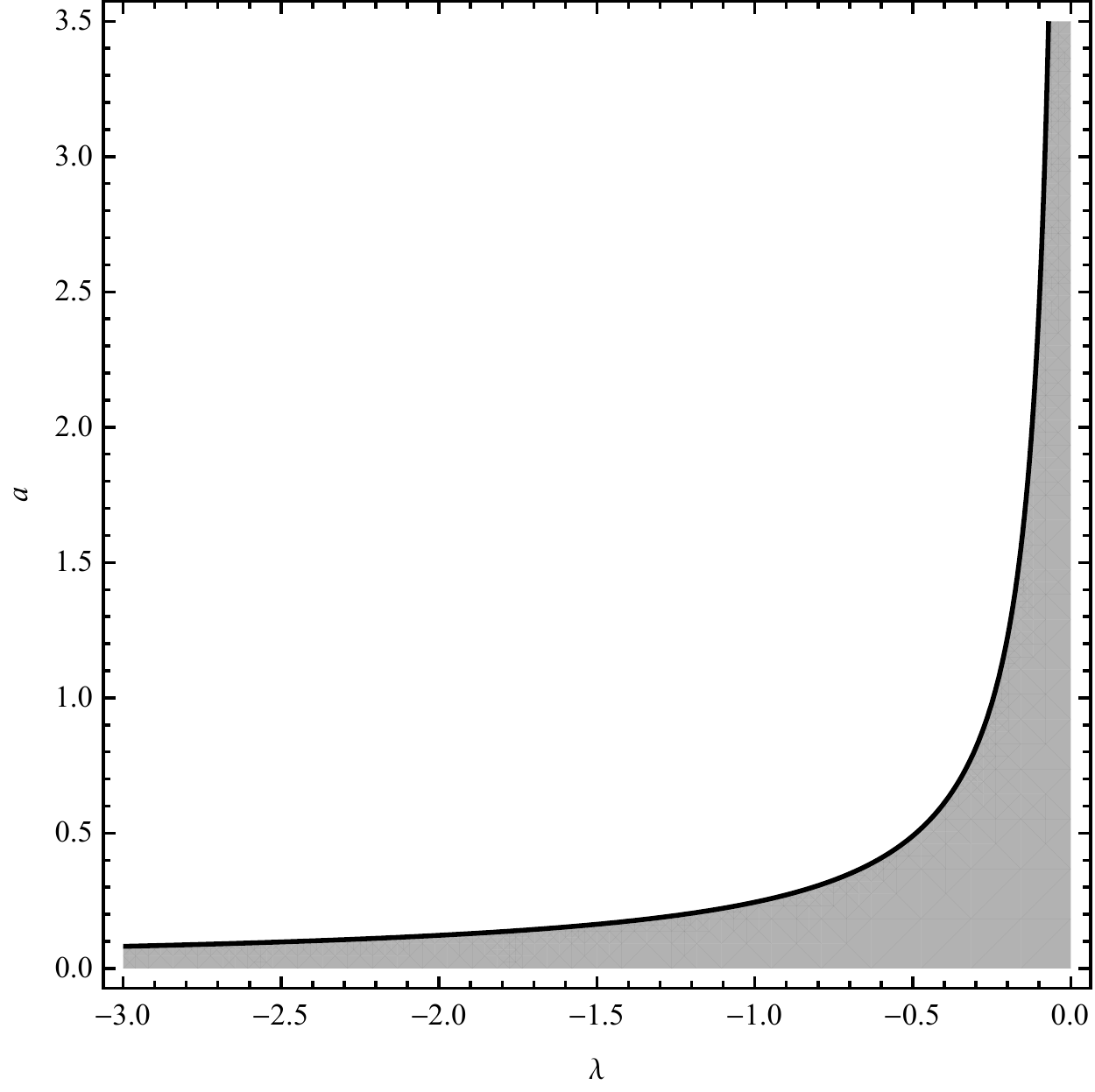}
\caption{$n=4$, the case for $\Lambda<0$ and $K=0$. In the shadow region, inequalities  (\ref{n4condition2})  and $1+4a\lambda>0$ are satisfied. }
\label{fig2}
\end{figure}

\item[(ii).] In the case with $1+4a\lambda=0$, we have $x\in\left(0,1\right]$, and then $y\in\left(0,1\right]$. $Q$ has a same form as in Eq.(\ref{Q5}).
It is obvious that  $Q=0$ when $n=3,5$,  and $Q<0$ when $n=4$, and $Q>0$ when $n\ge6$.

\item[(iii).] In the case  $1+4a\lambda<0$, we have 
$$x\in\left(\left(1+\frac{1}{4a\lambda}\right)^{\frac{1}{n+1}},1\right]\, ,$$
and then $y\in\left(0,1\right]$. 
From the expression (\ref{hatQ13}) and the condition $1+4a\lambda<0$, we have
we have $\hat{Q}(y)<0$ for $n=3,4,5$. Of course, this also means $Q<0$ for $n\le 5$.

For $n\ge6$, $\hat{Q}(y)$ has only one zero point $s$ in $(0,+\infty )$, where $s$ is given by Eq.(\ref{s1}). 
Therefore, when $1<s$, we have $Q<0$ in $(0,1]$; when $1>s$, we have $Q<0$ in $(0,s)$ and $Q>0$ in $(s,1]$.
\end{itemize}

\subsubsection{$\Lambda<0$ and $K=-1$ case}

For $n\ge3$, when
\begin{eqnarray}
\label{Mmin1}
\tilde{M}-\frac{1}{\tilde{\Lambda}(n+1)^2} \Bigg\{2^{\frac{3}{2}-\frac{n}{2}}\Bigg[-\frac{(n-1)+\sqrt{(n-1)^2+4\tilde{\alpha}\tilde{\Lambda}(n-3)(n+1)}}{(n+1)\tilde{\Lambda}}\Bigg]^{\frac{n-3}{2}}\nonumber\\
\times \left[-1+4\tilde{\alpha}\tilde{\Lambda}+n+4n\tilde{\alpha}\tilde{\Lambda}+\sqrt{(n-1)^2+4\tilde{\alpha}\tilde{\Lambda}(n-3)(n+1)}\right]\Bigg\}>0,
\end{eqnarray} 
%
there is a black hole horizon~\cite{Torii:2005xu}. With this condition, the outer event horizon is not degenerate.  Furthermore, to ensure the existence of the event horizon, we also have  $r_+> \sqrt{2\tilde{\alpha}}$, i.e., $a<1/2$. Since the mass parameter $\tilde{M}$ is not necessary to be positive when $K=-1$, we will discuss the hyperbolicity as follows.

(1): Firstly, let us consider the case with positive mass parameter $\tilde{M}$, i.e., the case with $\tilde{M}>0$. $\tilde{M}>0$ implies $a>1+\lambda$. 
So we get $$1+4a\lambda<(1-2a)^2\, .$$ 
The quantity $Q$ and its numerator $\hat{Q}$ still have forms (\ref{Q13}) and (\ref{hatQ13}), but now $y$ is defined as
$$y=\sqrt{1+4a\lambda+4a (-\lambda-1+a) x^{n+1}}\, .$$ 
\begin{itemize}
\item[(i).] If $1+4a\lambda>0$, then $y\in\left[\sqrt{1+4a\lambda},1-2a\right]$.
For $n=3$ and $n=5$, with the same logic, we have $\hat{Q}>0$, and then $Q>0$.
For $n=4$, we have
\begin{eqnarray}
	\hat{Q}(y)=-y^4+50(1+4a\lambda)y^2-25(1+4a\lambda)^2\, ,
\end{eqnarray}
and then 
\[\begin{cases}
\hat{Q}> 0\, ,& y\in\left[\sqrt{1+4a\lambda}\, ,\sqrt{1+4a\lambda}\cdot\sqrt{25+10\sqrt{6}}\right)\, ,\\
\hat{Q}<0\, , & y\in \left(\sqrt{1+4a\lambda}\cdot\sqrt{25+10\sqrt{6}}\, ,+\infty\right)\, .
\end{cases}\]
Therefore, $\hat{Q}>0$ for all $y$  in $[\sqrt{1+4a\lambda},1-2a]$ when 
\begin{equation}
\label{n4condition3}
1-2a<\sqrt{1+4a\lambda}\cdot\sqrt{25+10\sqrt{6}}\, .
\end{equation}
We also have $\hat{Q}<0$ for $y$ in $(\sqrt{1+4a\lambda}\cdot\sqrt{25+10\sqrt{6}},1-2a]$ when $$1-2a>\sqrt{1+4a\lambda}\cdot\sqrt{25+10\sqrt{6}}\, ,$$
and $\hat{Q}>0$ for $y$ in $[\sqrt{1+4a\lambda}\, ,\sqrt{1+4a\lambda}\cdot \sqrt{25+10\sqrt{6}})$.
%
For $n\ge6$, we find that 
$\hat{Q}(y)$ has only one  zero point $(\sqrt{1+4a\lambda})t$ in $\left(0,+\infty\right)$. Noted that $t<1$, we have $(\sqrt{1+4a\lambda})t<\sqrt{1+4a\lambda}$. So, we have $Q>0$.

So the $D=6$ is special. To ensure $Q>0$, the allowed parameters have been given in the shadow region of Fig.\ref{fig3}. The dashed line in Fig.\ref{fig3} is for $a=1+\lambda$. So the region above this line corresponds
to $\tilde{M}>0$. 

\begin{figure}[htbp]
\centering
\includegraphics[width=3.0 in]{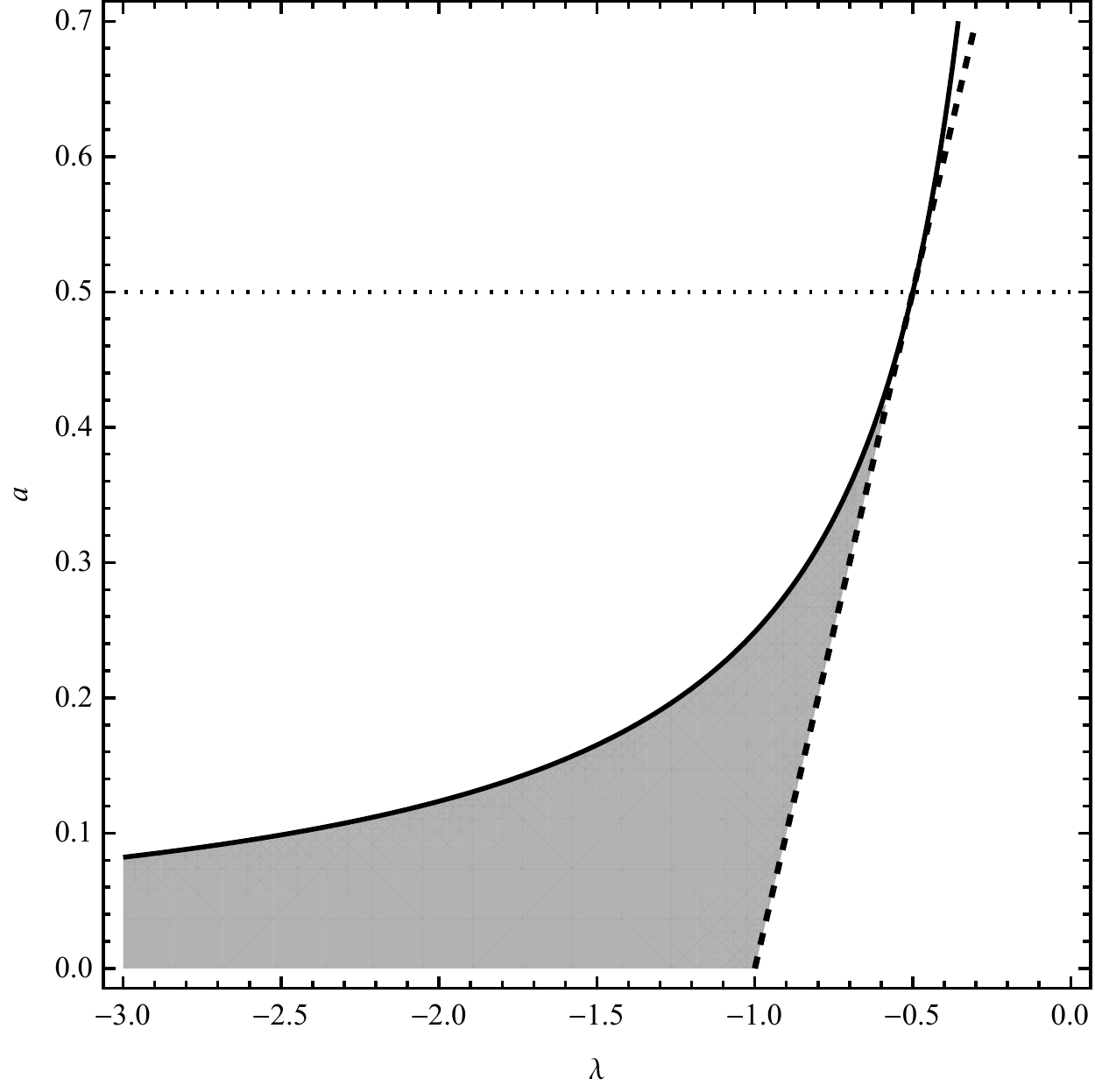}
\caption{$n=4$, the case for $\Lambda<0$ and $K=-1$. In the shadow region, inequalities  (\ref{n4condition3}) ,  $1+4a\lambda>0$,  $a>1+\lambda$, and $a<1/2$ are satisfied.  Above the dashed line, the mass parameter $\tilde{M}$ is positive. Bellow the solid black line, the hyperbolicity is satisfied outside the event horizon.}
\label{fig3}
\end{figure}

\item[(ii).] When $1+4a\lambda=0$, $x\in\left(0,1\right]$, and then $y\in\left(0,1-2a\right]$. Now $Q$ has a form in Eq.(\ref{Q5}). 
It is obvious that $Q=0$ when $n=3,5$, and $Q<0$ when $n=4$, and $Q>0$ when $n\ge6$.

\item[(iii).] When $1+4a\lambda<0$, we have 
$$x\in\left(\left[\frac{-(1+4a\lambda)}{4a(-\lambda-1+a)}\right]^{\frac{1}{n+1}},1\right]\, ,$$
and then $y\in\left(0,1\right]$. 
Obviously, we have $\hat{Q}(y)<0$ when $n=3\, ,4\, ,5$.  For $n\ge6$, $\hat{Q}(y)$ has only one zero point $s$ in $\left(0,+\infty\right)$, where $s$ is given by Eq.(\ref{s1}).
Therefore,  we have $Q<0$ for  all $y$ in $\left(0,1\right]$ when $1-2a<s$.  When $1-2a>s$, we have $Q<0$ for $y$ in $(0,s)$ and $Q>0$ for $y$ in $\left(s,1-2a\right]$. 
\end{itemize}

(2): In the case with $\tilde{M}=0$. Noted that $0\le1+4a\lambda<1$, we have
\begin{equation}
	Q=\sqrt{1+4a\lambda}\ge0\, .
\end{equation}

(3): In the case with $\tilde{M}<0$. Obviously, we have $0<1+4a\lambda<1$. 
$\tilde{M}<0$ implies $a<1+\lambda $, and we have $$\sqrt{1+4a\lambda}>1-2a\, .$$
Consider the region outside the event horizon, i.e., the region with $x\in\left[0,1\right]$, then we can define
$$y=\sqrt{1+4a\lambda+4a(-\lambda-1+a)x^{n+1}}\, .$$
It is easy to find  $y\in\left[1-2a,\sqrt{1+4a\lambda}\right]$, and $\hat{Q}$ has the same form as the one in (\ref{hatQ13}).  When $K=-1$, the black hole might have
inner horizon~\cite{Torii:2005xu}. Inner and outer event horizons both exist when
\begin{eqnarray}
\label{innerouter1}
	(2\tilde{\alpha})^{\frac{n-3}{2}}\tilde{\alpha}(1+4\tilde{\alpha}\tilde{\Lambda})+\tilde{M}<0\, ,
\end{eqnarray}
and only one event horizon exists when
\begin{eqnarray}
\label{innerouter2}	
(2\tilde{\alpha})^{\frac{n-3}{2}}\tilde{\alpha}(1+4\tilde{\alpha}\tilde{\Lambda})+\tilde{M}>0\, .
\end{eqnarray}
For $n=3$, we have
\begin{eqnarray}
		\hat{Q}(y)=24(1+4a\lambda)y^2-16(1+4a\lambda)^2\nonumber\, .
\end{eqnarray}
This implies
\[\begin{cases}
\hat{Q}> 0\, ,& y\in\left( \sqrt{2(1+4a\lambda)/3}\, ,\sqrt{1+4a\lambda}\right]\, ,\\
\hat{Q}<0\, , & y\in \left(0\, ,\sqrt{2(1+4a\lambda)/3}\right)\, .
\end{cases}\]
Thus, we have $\hat{Q}>0$ for all $y$ in $[1-2a,\sqrt{1+4a\lambda}]$when $$1-2a>\sqrt{2(1+4a\lambda)/3}\, .$$
We have $\hat{Q}<0$ for $y$ in $[1-2a,\sqrt{2(1+4a\lambda)/3})$ when $$1-2a<\sqrt{2(1+4a\lambda)/3}\, ,$$
and $\hat{Q}>0$ for $y$ in $(\sqrt{2(1+4a\lambda)/3},\sqrt{1+4a\lambda}]$.

For $n=4$, we have
\begin{eqnarray}
\hat{Q}(y)=-y^4+50(1+4a\lambda)y^2-25(1+4a\lambda)^2\nonumber\, .
\end{eqnarray}
This equation gives
\[\begin{cases}
\hat{Q}> 0\, ,& y\in \left(\sqrt{1+4a\lambda}\cdot\sqrt{25-10\sqrt{6}},\sqrt{1+4a\lambda}\right]\, ,\\
\hat{Q}<0\, , & y\in \left(0,\sqrt{1+4a\lambda}\cdot\sqrt{25-10\sqrt{6}}\right)\, .
\end{cases}\]
So we have $\hat{Q}>0$ for all $y$ in $[1-2a,\sqrt{1+4a\lambda}]$ when 
\begin{equation}
\label{n4condition51}
1-2a>\sqrt{1+4a\lambda}\cdot\sqrt{25-10\sqrt{6}}\, .
\end{equation}
However, we have $\hat{Q}<0$ for $y$  in $[1-2a,\sqrt{1+4a\lambda}\cdot\sqrt{25-10\sqrt{6}})$ when $$1-2a<\sqrt{1+4a\lambda}\cdot\sqrt{25-10\sqrt{6}}\, ,$$and $\hat{Q}>0$ for $y$ in $(\sqrt{1+4a\lambda}\cdot\sqrt{25-10\sqrt{6}},\sqrt{1+4a\lambda}]$.

For $n=5$, we have
\begin{eqnarray}
	\hat{Q}(y)=84(1+4a\lambda)y^2-36(1+4a\lambda)^2\nonumber\, .
\end{eqnarray}
This equation tells us we have 
\[\begin{cases}
\hat{Q}> 0\, ,& y\in \left(\sqrt{3(1+4a\lambda)/7},\sqrt{1+4a\lambda}\right]\, ,\\
\hat{Q}<0\, , & y\in \left(0,\sqrt{3(1+4a\lambda)/7}\right)\, .
\end{cases}\]
So we have $\hat{Q}>0$ for all $y$ in $\left[1-2a,\sqrt{1+4a\lambda}\right]$
 when $$1-2a>\sqrt{3(1+4a\lambda)/7}\, .$$
We have $\hat{Q}<0$ for $y$ in $[1-2a,\sqrt{3(1+4a\lambda)/7})$ when 
  $$1-2a<\sqrt{3(1+4a\lambda)/7}\, ,$$ 
  and $\hat{Q}>0$ for $y$ in $(\sqrt{3(1+4a\lambda)/7},\sqrt{1+4a\lambda}]$.
     
For $n\ge6$, we have
\begin{equation}
	\hat{Q}(y)=(n-3)(n-5)y^4+2(n+1)(2n-3)(1+4a\lambda)y^2-(n+1)^2(1+4a\lambda)^2\nonumber\, .
\end{equation}
It is obvious that $\hat{Q}(y)$ has only one  zero point $(\sqrt{1+4a\lambda}) t$ in $\left(0,+\infty\right)$, where $t$ is given by Eq.(\ref{t1}).
Noted  that $t<1$, we have $(\sqrt{1+4a\lambda}) t<\sqrt{1+4a\lambda}$. Therefore, we have  $\hat{Q}>0$ for $y$ in $\left[1-2a,\sqrt{1+4a\lambda}\right]$
when $$1-2a>(\sqrt{1+4a\lambda})t\, .$$
We also have $\hat{Q}<0$ for $y$ in $\left[1-2a,\sqrt{1+4a\lambda}t\right)$  when $$1-2a<(\sqrt{1+4a\lambda})t\, ,$$ and $\hat{Q}>0$ for $y$ in $\left((\sqrt{1+4a\lambda})t,\sqrt{1+4a\lambda}\right]$.

\begin{figure}[htbp]
\centering
\includegraphics[width=3.0 in]{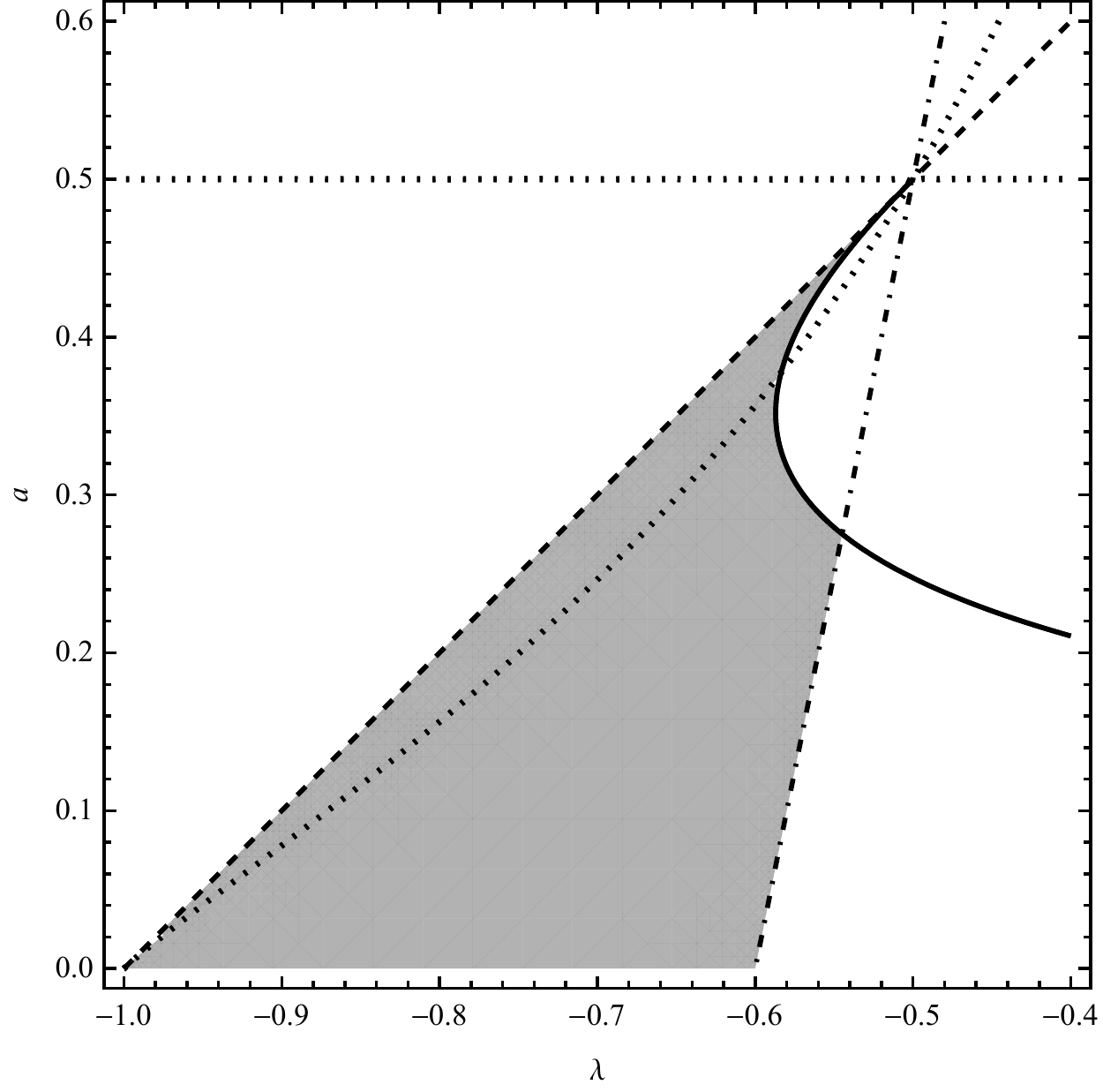}
\caption{$n=4$, the case for $\Lambda<0$ and $K=-1$.  In the shadow region, inequalities (\ref{Mmin1}), (\ref{n4condition51}), $a>5\lambda +3$, and $a<1+\lambda$ are satisfied. Above the dotted line in the shadow region, the black hole has only one horizon. Bellow
this dotted line, the black hole has two horizons.  The dotdashed line is for $a > 5\lambda + 3$. The solid black line is for the hyperbolicity. }
\label{fig5}
\end{figure}

So the situation is complicated when $\tilde{M}<0$. In this case, hyperbolicity can be broken outside the event horizon for all dimension $D\ge 5$. Here, as an example,  Fig.\ref{fig5} for $D=6$ or $n=4$ has been given to show the details of the range of the parameters. The dotted line in the shadow region of Fig.\ref{fig5}  is for the conditions (\ref{innerouter1}) and  (\ref{innerouter2}).  When condition (\ref{innerouter1}) is satisfied, the solution has two event horizon. To ensure $r_+$ is the radius of outer event horizon, we have to impose a
condition $F_x(x)<0$ at $x=1$ (noted that $x=r_+/r$), and this gives 
$$a > 5\lambda + 3\, .$$
This corresponds to the dotdashed line in Fig.\ref{fig5}.

\subsubsection{$\Lambda>0$ and $K=1$ case}

Although cosmological horizon  exists when $\Lambda>0$, the discussion on the hyperbolicity is similar to the case with $K=1$ and $\Lambda<0 $.  Maybe the most significant difference is the existence conditions for the black hole solutions.

For $n=3$, when $\tilde{M}>\tilde{\alpha}$ and $$1+4\tilde{\alpha}\tilde{\Lambda}-4\tilde{\alpha}\tilde{M}>0\, ,$$ there is a black horizon~. For $n\ge 4$
\begin{eqnarray}
\label{Mmax}
\tilde{M}-\frac{1}{\tilde{\Lambda}(n+1)^2} \Bigg\{2^{\frac{3}{2}-\frac{n}{2}}\Bigg[\frac{(n-1)+\sqrt{(n-1)^2+4\tilde{\alpha}\tilde{\Lambda}(n-3)(n+1)}}{(n+1)\tilde{\Lambda}}\Bigg]^{\frac{n-3}{2}}\nonumber\\
\times \left[-1+4\tilde{\alpha}\tilde{\Lambda}+n+4n\tilde{\alpha}\tilde{\Lambda}+\sqrt{(n-1)^2+4\tilde{\alpha}\tilde{\Lambda}(n-3)(n+1)}\right]\Bigg\}<0,
\end{eqnarray} 
with $\tilde{M}>0$,  black hole solution exists~\cite{Torii:2005xu}.  The above inequality suggests  a bound on the  black hole, i.e., the radius of black hole horizon must be less than the radius of cosmological horizon.
Outside the event horizon of this solution, we have $x\in\left[0,1\right]$. let 
$$y=[1+4a\lambda+4a(-\lambda+1+a)x^{n+1}]^{\frac{1}{2}}\, ,$$
then  $y\in\left[\sqrt{1+4a\lambda},1+2a\right]$.  By this definition, the numerator of $Q$ is the same as the one in Eq.(\ref{Q13}). Except  $\Lambda>0$,  all of the discussion and results are the same as the case with $K=1$ and $\Lambda<0 $.

For $n=3$, $n=5$, and $n\ge 6$, $Q$ is always positive outside the black hole horizon.  Actually, $Q$ is positive for $x\in [0\, ,+\infty)$, where $+\infty $ corresponds to $r=0$, i.e.,  the singularities of the spacetimes.

$D=6$ or $n=4$ is also special, the numerator of  $Q$ is still given by (\ref{Q41}), so we have
%
\[\begin{cases}
\hat{Q}> 0\, ,& y\in\left[\sqrt{1+4a\lambda},\sqrt{1+4a\lambda}\cdot\sqrt{25+10\sqrt{6}}\right)\, ,\\
\hat{Q}<0\, , & y\in \left(\sqrt{1+4a\lambda}\cdot\sqrt{25+10\sqrt{6}},+\infty\right)\, .
\end{cases}\]
Thus, we have $Q>0$ for all  $y\in \left[\sqrt{1+4a\lambda},1+2a\right] $ when 
\begin{equation}
\label{n4conditionP1}
1+2a<\sqrt{1+4a\lambda}\cdot\sqrt{25+10\sqrt{6}}\, .
\end{equation}
However, we have $Q<0$ if $y\in \big(\sqrt{1+4a\lambda}\cdot\sqrt{25+10\sqrt{6}},1+2a\big]$ when
$$1+2a>\sqrt{1+4a\lambda}\cdot\sqrt{25+10\sqrt{6}}\, ,$$
and $Q>0$ if $y\in [\sqrt{1+4a\lambda}\, , \sqrt{1+4a\lambda}\cdot \sqrt{25+10\sqrt{6}})$. 

\begin{figure}[htbp]
\centering
\includegraphics[width=3.0 in]{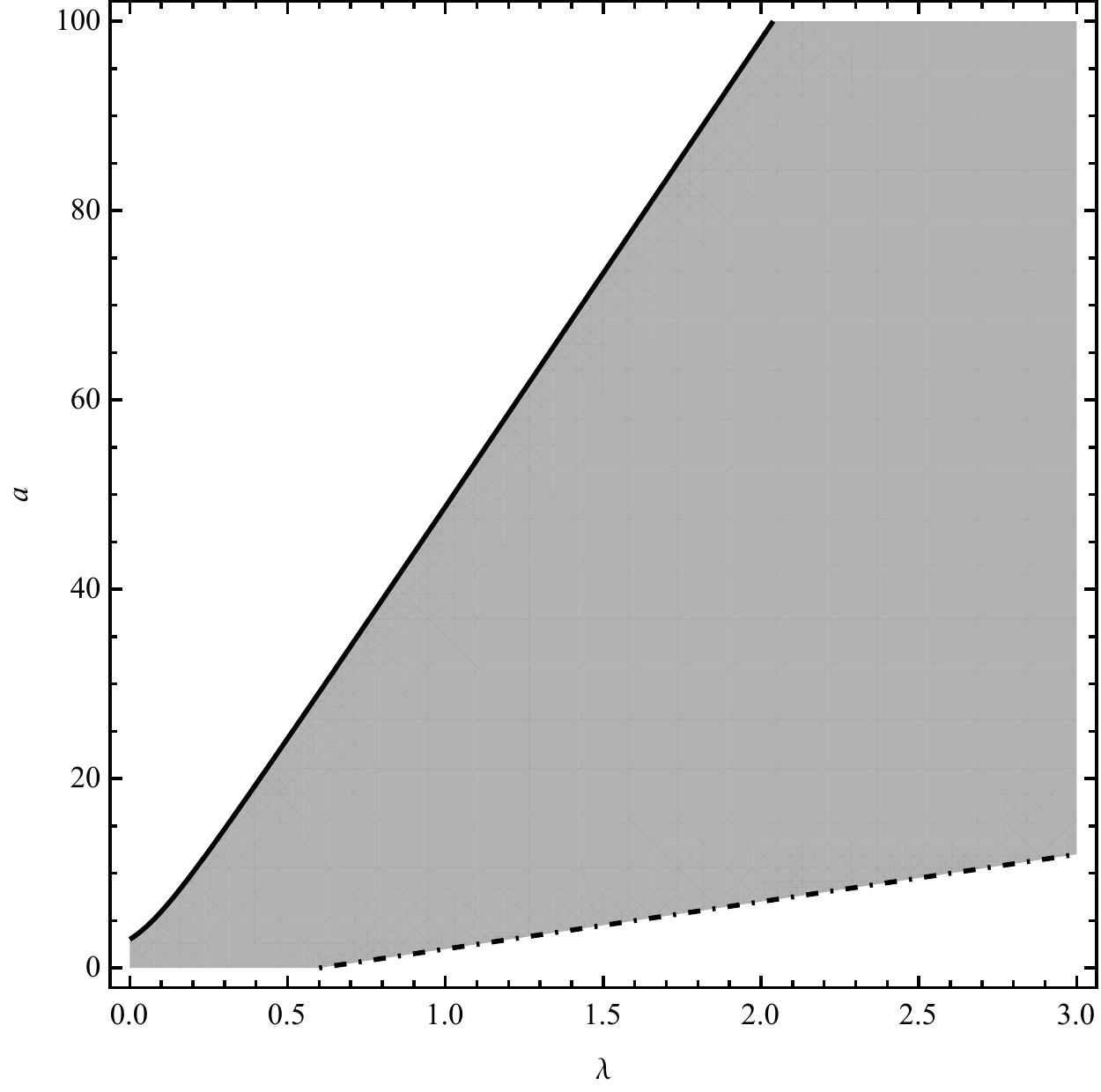}
\caption{$n=4$, the case for $\Lambda>0$ and $K=1$. In the shadow region, inequalities (\ref{Mmax}),  (\ref{n4conditionP1}),  and   $a>5\lambda -3 $ are satisfied. }
\label{fig4}
\end{figure}

It should be point out here:  There are two horizons in this case, i.e., black hole horizon and cosmological horizon. The parameter $x$ is defined by the 
radius of black hole horizon, i.e., $x=r_+/r$, so we have $F_x(x)<0$ at $x=1$. For $n=4$, $F_x(x)<0$ at $x=1$ gives a constraint
\begin{equation}
\label{alambda1}
a > 5\lambda -3\, .
\end{equation}
Above discussion show that $D=6$ is special, we have to impose a condition (\ref{n4conditionP1}) on the parameter $a$ and $\lambda$ to ensure the positiveness of $Q$ outside the event horizon. 
The allowed range of the parameters have been depicted in Fig.\ref{fig4}.  The black solid line in Fig.\ref{fig4} is for the condition (\ref{n4conditionP1}). The region below this line 
has a positive $Q$. The dotdashed line in Fig.\ref{fig4} is for (\ref{alambda1}). 

\subsection{regular solution}
A globally regular solution of the theory  only happens when $\tilde{M}=0$ (The case with $M=0$, $K=-1$, and $\Lambda<0$ is an exception, which has a horizon and  has been discussed in the previous part of the paper).     
For $\tilde{M}=0$, we have
\begin{equation}
	Q=\sqrt{1+\frac{8(n-1)(n-2)\alpha\Lambda}{n(n+1)}}\, .
\end{equation} 
So, for the regular solution, we always have $Q\ge 0$. Of course, the above result also implies 
\begin{equation}
	1+\frac{8(n-1)(n-2)\alpha\Lambda}{n(n+1)}\ge 0\, .
\end{equation} 
For $\Lambda\ge 0$, this inequality is obviously right. However, for $\Lambda<0$, the Gauss-Bonnet coupling $\alpha$ has an upper bound which can be read out from the above immediately.

\subsection{The existence of superluminal mode}

When the tensor perturbation equation is hyperbolic, there are travel modes in the theory, and we can discuss the speed of a gravitational fluctuation.  In this subsection, we study the causality of this kind of tensor perturbation.

We investigate the possible superluminal modes by solving the inequalities (\ref{IEQ1}-\ref{IEQ3}). Actually from these inequalities,  we can get the precise conditions for the existence of a superluminal mode. For the black hole solutions of the theory, we find that   the superluminal mode is allowed only in the case with $\tilde{M}>0$. Further,  the superluminal mode could exist near the infinity of the spacetime. These can be found as follows.

Now, we find that $I$ and $\hat{J}$ can be expressed as
\begin{equation}
	I=\frac{2a(n-3)(K+aK^2-\lambda)x^{n+1}+(n-1)(1+4a\lambda)}{(n-1)[1+4a\lambda+4a(K+aK^2-\lambda)x^{n+1}]^{\frac{1}{2}}}\, ,
\end{equation}
and
\begin{equation}
	\hat{J}=\frac{2(n+1)(K+aK^2-\lambda)x^{n-1}\left[-a(n-3)(K+aK^2-\lambda)x^{n+1}+1+4a\lambda\right]}{\left[1+4a\lambda+4a(K+aK^2-\lambda)x^{n+1}\right]^{\frac{3}{2}}}\, .
\end{equation}
Firstly, we consider the cases with  $1+4a\lambda>0$. With this condition,  we  discuss  $n=3$, $n\ge 4$ separately.
\begin{itemize}
\item[(i).]
For $n=3$, we have $I>0$.
When $\tilde{M}>0$, i.e., $K+aK^2-\lambda>0$, we have $\hat{J}>0$. Therefore, there is always superluminal mode outside the black hole horizon if $\tilde{M}>0$. 
When $\tilde{M}\le 0 $, i.e., $K+aK^2-\lambda\le0$, we always have $\hat{J}\le0$. So the superluminal mode is absent when $\tilde{M}<0$.

\item[(ii).]For $n\ge4$, 
when $\tilde{M}>0$, i.e., $K+aK^2-\lambda>0$, we have $I>0$. By solving $\hat{J}>0$, we obtain 
\begin{eqnarray}
\label{rangex1}
	0\le x<\left[\frac{1+4a\lambda}{a(n-3)(K+aK^2-\lambda)}\right]^{\frac{1}{n+1}}\, .
\end{eqnarray}
This is the condition for the existence of superluminal mode. When $\tilde{M}=0$, i.e.,  $K+aK^2-\lambda=0$, the superluminal mode is absent. Actually, in this case, we have $I>0$ and $\hat{J}=0$. When $\tilde{M}<0$, i.e., $K+aK^2-\lambda<0$, we always have $\hat{J}<0$ no matter whether $I$ is positive  or not. Therefore, there is no superluminal mode when $\tilde{M}<0$.
\end{itemize}

In the case with $1+4a\lambda\le 0$, we always have $K+aK^2-\lambda\ge 0$.  This suggests that $\hat{J}\le0$. So there is no superluminal mode in this case.

\vspace{0.1cm}

At the end of this section, based on the exact range in (\ref{rangex1}), we list the detailed conditions for the existence of superluminal modes when $K=1$, $\lambda=0$, $a=24$. When $r > r_c$, where $r_c$ is determined by the rightmost term in the inequality  (\ref{rangex1}),  there will be superluminal mode. This result is consistence with the one in~\cite{Reall:2014pwa}.
\begin{center}
 \linespread{1.5}
 \begin{tabular}{|c|c|}
  \hline
  Dimensions & critical value $r_c$ ($r_+=1$) \\
  \hline
  $D=6\, ,n=4$ & 3.5944 \\
  \hline
  $D=7\, ,n=5$ & 3.2598 \\
  \hline
  $D=8\, ,n=6$& 2.9177 \\
  \hline
  $D=9\, ,n=7$& 2.6456 \\
  \hline
 \end{tabular}
\end{center}


\section{Nariai-type spacetime}
\label{section5}
When $r=r_0=\text{constant}$, equations of motion imply that ${}^2\!{R}$ is a constant. It means that $(M^2,g_{ab})$ is a two-dimensional constant curvature spacetime. This kind solution is called the Nariai-type spacetime~\cite{Maeda:2007uu}, and the metric in the standard coordinates is given by
\begin{equation}
\label{Nariai1}
	ds^2=-(1-\sigma\rho^2)dt^2+\frac{d\rho^2}{1-\sigma\rho^2}+r_0^2\gamma_{ij}dz^idz^j\, ,
\end{equation}
where 
\begin{equation}
	\sigma=\left[\frac{(n-1)+2(n-1)(n-2)(n-3)\alpha Kr_0^{-2}}{r_0^2+2(n-1)(n-2)\alpha K}\right]K\, .
\end{equation}
From Eq.$(\ref{EOM_3})$, we know that $r_0^2$ is the real and positive root of the following algebraic equation
\begin{equation}\label{algebraic}
	2\Lambda=\frac{n(n-1)K}{r_0^2}+\frac{n(n-1)(n-2)(n-3)\alpha K^2}{r_0^4}\, .
\end{equation}
 Of course, this equation does not always exist a real and positive root $r_0^2$. The simplest case is the solution with $K=0$. Obviously, $\Lambda$ and $\sigma$ have to be vanishing when $K=0$, and $r_0^2$ is an arbitrary  positive constant. So the solution (\ref{Nariai1}) reduces
 to a very simple form. In the case with $K\ne 0$, 
 the condition for real and positive $r_0^2$ depends on the dimension $n$.  For $n=3$, this condition is  $K\Lambda>0$.  For $n\ge4$, the condition becomes ~\cite{Maeda:2007uu}
  \[\begin{cases}
\Lambda > 0\, ,& K=\pm 1\, ,\\
\Lambda=0\, , &  K=-1\, ,\\
-n(n-1)/\left[8(n-2)(n-3)\alpha\right] < \Lambda <0\, , & K=-1\, .
\end{cases}\]
From the metric (\ref{Nariai1}), it is easy to find
 $${}^2\!{R}=2\sigma\, ,\quad  \prescript{2}{}{\Box}r=0\, ,\quad (Dr)^2=0\, .$$
 Therefore, $P$, $Q$,  $I$, and $J$ have following forms 
\begin{equation}
\label{PN1}
	P=-\left[1+2\alpha(n-2)(n-3)\frac{K}{r_0^2}\right]^2\, ,
\end{equation}
\begin{equation}
\label{QN1}
	Q=\frac{r_0^4+4(n^2-4n+6)Kr_0^2\alpha+4(n-1)(n-2)^2(n-3)K^2\alpha^2}{r_0^2[r_0^2+2\alpha(n-1)(n-2)K]}\, .
\end{equation}
\begin{equation}
\label{IN1}
	I=1+2\alpha (n-2)(n-3)\frac{K}{r_0^2}\, ,
\end{equation}
and
\begin{equation}
\label{JN1}
	J=\frac{4 K}{r_0^2+2(n-1)(n-2)\alpha}\, .
\end{equation}
Obviously, $P$ can be vanishing only in the case with negative $K$.  Roughly speaking, we find that the superluminal modes appear in the cases  $K=\pm 1$   with some special $\alpha$ and $\Lambda$. Details can be found in the following subsections.

\subsection{$K=1$ case}
From the expression (\ref{PN1}), one has $P<0$. It is also easy to find  that (\ref{QN1}) implies
\begin{equation}
	Q=\frac{r_0^4+4(n^2-4n+6)r_0^2\alpha+4(n-1)(n-2)^2(n-3)\alpha^2}{r_0^2[r_0^2+2\alpha(n-1)(n-2)]}>0\, ,
\end{equation}
It is easy to find that $I$ and $J$ are both positive when $K=1$. 
Therefore, there is always superluminal modes in this case.

\subsection{$K=0$ case} 
When $K=0$, we have $\sigma=0$ and $\Lambda=0$. The metric can be represented as 
\begin{equation}
	ds^2=-dt^2+d\rho^2+r_0^2\delta_{ij}dz^idz^j.
\end{equation}
In this case, $P$, $Q$, $I$, $J$ have  very simple forms i.e., $P=-1$, $Q=1$, $I=1$, $J=0$. Therefore, there is no superluminal modes in this case.

\subsection{$K=-1$ case}
This case is a little bit complicated. However, Eq.(\ref{PN1}) implies that $P$ is always nonpositive. Now $Q$ has a form
\begin{equation}
	Q=\frac{r_0^4-4(n^2-4n+6)r_0^2\alpha+4(n-1)(n-2)^2(n-3)\alpha^2}{r_0^2[r_0^2-2\alpha(n-1)(n-2)]}\, .
\end{equation}
So $Q$ might be negative.  From the expressions (\ref{IN1}) and (\ref{JN1}),  
we find that the signs of  $I$ and $J$  depend on the parameters and dimensions.  We will discuss the cases with $\Lambda=0$, $\Lambda>0$, and $\Lambda<0$ separately.

\subsubsection{$\Lambda=0$ case}
In the case with $n=3$, there is a contradiction in Eq.(\ref{algebraic}). For $n\ge4$, we have $r_0^2=(n-2)(n-3)\alpha$. In this case, 
\begin{equation}
	Q=-\frac{n^2-n-10}{(n+1)(n-2)}<0.
\end{equation}
Since $Q<0$ in this case, the discussion of the causality is meaningless.

\subsubsection{$\Lambda>0$ case}
In this case, Eq.(\ref{algebraic}) gives
\begin{equation}
	r_0^2=\frac{n}{4\Lambda}\Bigg\{-(n-1)+\sqrt{(n-1)^2+\frac{8\Lambda}{n}(n-1)(n-2)(n-3)\alpha}\Bigg\}\, .
\end{equation}
It is not hard to find that there is no $\alpha$ and $\Lambda$ satisfying the condition $P<0$ and $Q>0$. So $Q$ is always negative, the tensor perturbation equation is not hyperbolic.

\subsubsection{$\Lambda<0$ case}
In this case, Eq.(\ref{algebraic}) tells us
\begin{equation}
\label{r02}
	r_0^2=\frac{n}{4\Lambda}\Bigg\{-(n-1) \pm \sqrt{(n-1)^2+\frac{8\Lambda}{n}(n-1)(n-2)(n-3)\alpha}\Bigg\}\, .
\end{equation}
Solving the condition $P<0$, $Q>0$, $I>0$ and $J>0$, we have the following results:

\begin{center}
\linespread{1.5}
\begin{tabular}{|c|c|c|}
	\hline
	dimenion & hyperbolicity condition & causality condition\\
	\hline
	$n=3$ & $\displaystyle\alpha\Lambda<-\frac{3}{4}$ or $\displaystyle-\frac{1}{4}<\alpha\Lambda<0$ &  $\displaystyle-\frac{1}{4}<\alpha\Lambda<0$ \\
	\hline
	$n\ge4$ $(+)$ & $\displaystyle-\frac{n(n-1)}{8(n-2)(n-3)}<\alpha\Lambda<M(n)$ & $\displaystyle-\frac{n(n-1)}{8(n-2)(n-3)}<\alpha\Lambda<M(n)$\\
	\hline
	$n \ge 4 $ $(-)$& $\displaystyle-\frac{n(n-1)}{8(n-2)(n-3)}<\alpha\Lambda<-\frac{n(n+1)}{8(n-2)(n-1)}$ or $\displaystyle N(n)<\alpha\Lambda<0$&$N(n)<\alpha\Lambda<0$\\
	\hline
\end{tabular}
\end{center}
\vspace{0.1cm}
In the above table,  $(+)$ and $(-)$ correspond to the sign $\pm$ in (\ref{r02}), and
\begin{equation}
	M(n)=\frac{-n(n^4-6n^3+7n^2+16n-36)-2n(n-4)\sqrt{5n^2-20n+24}}{8(n-1)(n-2)^3(n-3)}\, ,
\end{equation}
\begin{equation}
	N(n)=\frac{-n(n^4-6n^3+7n^2+16n-36)+2n(n-4)\sqrt{5n^2-20n+24}}{8(n-1)(n-2)^3(n-3)}\, .
\end{equation}
In the $``(+) "$  case of $n\ge 4$, it is easy to find that the range of $\alpha \Lambda$ become narrower and narrower when $n$ increases.  It is also not hard to find
that $M(n)$ and $N(n)$ are both monotonically increase when $n\ge 5$. Therefore,  if $$-\frac{1}{8} < \alpha \Lambda < 0\, ,$$the hyperbolicity and causality will be satisfied in all dimensions.


\section{Dimensionally extended constant curvature black hole}
\label{section6}

Consider the equations of motion, especially the Eq.(\ref{EOM_2}),   when  
$$1+ 2 \tilde{\alpha} \frac{K - (Dr)^2}{r^2}=0\, ,$$ one gets a solution with $1+4\tilde{\alpha}\tilde{\Lambda}=0$. This corresponds to the dimensionally extended constant curvature black hole given by Banados, Teitelboim and Zanelli~\cite{Maeda:2007uu, Banados:1992gq}.  This kind of solution is given by 
 \begin{equation}
	ds^2=-h(r)e^{2\delta(t,r)}dt^2+h^{-1}(r)dr^2+r^2\gamma_{ij}dz^idz^j\, ,
\end{equation}
where 
\begin{equation}
\label{h1}
h(r)=K+\frac{r^2}{2\tilde{\alpha}}\, ,
\end{equation}
and $\delta(t,r)$ is an arbitrary function.
For this spacetime, we find that $P$ is always vanishing, and
\begin{equation}
\label{QN23}
	Q=-2\alpha\left[3h^{\prime}\delta^{\prime}+2h\delta^{\prime\prime}+2h^2\left(\delta^{\prime}\right)^2+2(n-3)\frac{h\delta^{\prime}}{r}\right]\, ,
\end{equation}
where $\prime$ stands for $\partial/\partial r$. Obviously, $Q$ is also vanishing if $\delta$ does not depend on $r$, i.e., $\delta=\delta(t)$.  So the principle symbol 
for the tensor perturbation equation in this background is totally degenerated.

Here we show why that $P$ is always vanishing.  From the solution, it is not hard to find
\begin{equation}
	D_aD_brD^aD^br=\frac{1}{2}\left[\left(h^{\prime}\right)^2+2hh^{\prime}\delta^{\prime}+2h^2\left(\delta^{\prime}\right)^2\right]\, .
\end{equation}
By using 
\begin{equation}
	\prescript{2}{}{\Box}r=h^{\prime}+h\delta^{\prime}\, ,
\end{equation}
we get
\begin{equation}
	D_aD_brD^aD^br=\frac{1}{2}\left[\left(\prescript{2}{}{\Box}r\right)^2+\frac{1}{2}h^2\left(\delta^{\prime}\right)^2\right]\, .
\end{equation}
Therefore, substituting this result into the expression (\ref{PAppendix}) in Appendix B, we obtain
\begin{eqnarray}
		P= \frac{4(n-2)^2}{r^2}\alpha^2h^2\left(\delta^{\prime}\right)^2-\left\lbrace1-2\alpha(n-2)\left[\frac{\prescript{2}{}{\Box}r}{r}-(n-3)\frac{K-(Dr)^2}{r^2}\right]\right\rbrace^2\, .
\end{eqnarray}
Consider that function $h$ is given by (\ref{h1}), we obtain $P=0$.  With the above relations, we also get the expression of $Q$, i.e.,  Eq.(\ref{QN23}).

The above calculation shows the tensor perturbation equation is not hyperbolic in this case, and  there is no gravitational wave on this dimensionally extended constant curvature black hole spacetime.


\section{Vaidya spacetimes}
\label{section7}
A Vaidya spacetime is a solution of the theory with radiation matter, the  metric of the spacetime can be expressed as
\begin{eqnarray}
	ds^2=-F(v\, ,r)dv^2+2dv dr+r^2\gamma_{ij}dz^idz^j\, .
\end{eqnarray} 
In general, the energy-momentum tensor of the radiation matter satisfy 
$$T_{\ell\ell}\ne0\, ,\quad \mathrm{and}\quad  T_{nn}=0\, ,$$ 
or 
$$T_{\ell\ell}=0\, ,\quad \mathrm{and}\quad  T_{nn}\ne 0\, .$$
Therefore, from Eq.(\ref{P_nomal}), we have $P\le 0$. From the above metric, we have
\begin{equation}\label{V_Q}
	{}^2\!{R}=-F^{\prime\prime},\quad\prescript{2}{}{\Box}r=F^{\prime},\quad(Dr)^2=F\, ,
\end{equation}
where $\prime$ stands for $\partial/\partial r$ as before. Substituting the quantities in (\ref{V_Q}) into Eq.(\ref{Q}), we can get
\begin{eqnarray}
	Q=1+2\alpha\left[-F^{\prime\prime}-2(n-3)\frac{F^{\prime}}{r}+(n-3)(n-4)\frac{K-F}{r^2}\right]\, ,
\end{eqnarray}
where 
\begin{eqnarray}
	F(v\, ,r)=K+\frac{r^2}{2\tilde{\alpha}}\left[1\mp\sqrt{1+\frac{8\kappa_{D}^2\tilde{\alpha}M(v)}{nV_{n}^Kr^{n+1}}+4\tilde{\alpha}\tilde{\Lambda}}\right]\, .
\end{eqnarray}
The trapping horizon or apparent horizon of Vaidya spacetime is given by $F(v,r)=0$, and we can get the radius of the apparent horizon $r_A(v)$.  Instead of the radius of event horizon $r_+$ in Eq.(\ref{xalambda}), in the untrapped region
of the spacetime, the radius of the apparent horizon $r_A(v)$ can be used 
to define $x$,  $a$, and $\lambda$, i.e., we have
$$x=\frac{r_A}{r}\, ,\quad a=\frac{\tilde{\alpha}}{r_A^2}\, ,\quad \lambda =r_A^2\tilde{\Lambda}\, .$$
Although these quantities depend on $v$, in the discussion of the hyperbolicity and causality, the algebraic structure is the same as  the case of  the static, and we can get nearly the same conclusions as the static cases. We will not repeat this kind of discussion here.


\section{Summary and discussion}
\label{section8}

In this paper, we have obtained  the general master equations of tensor type for the warped spacetimes with the metric (\ref{metric}), i.e., Eq.(\ref{PMNIJ1}) or Eq.(\ref{PE1}). These new master equations do not depend on
the mode expansion.  Of course, one can introduce the harmonic tensor $\mathds{T}_{ij}$ on the maximally symmetric  space $(N\, ,\gamma_{ij})$~\cite{Higuchi:1986wu},  i.e., the functions satisfying
$$ \hat{\Delta} \mathds{T}_{ij} = (-  k^2+ 2n K)\mathds{T}_{ij} \, ,$$
and  expand  $h_{ij}=h^{\mathrm{TT}}_{ij}$ as $h_{ij} /r^2= h_T \mathds{T}_{ij}$ (the summation on $k$ implied), then we can get the equation for each mode $h_{T}$. Actually, in Appendix A, we have get the more general master equations (\ref{master1}) for 
the $D=m+n$ dimension warped spacetimes in which $(N\, ,\gamma_{ij})$ is an Einstein manifold. 

 Based on the master equation, especially the effective potential in~\cite{Takahashi:2010ye}, Reall has provided a smart way to study the hyperbolicity and causality of the perturbation equations in Einstein-Gauss-Bonnet theory~\cite{Reall:2014pwa}. 
 This method is focused  on the large $k$ limit in the mode expansion. Now, from our formula (\ref{PE1}),  it is obvious this limit can help people extract the information about $P^{ij}\hat{D}_i\hat{D}_j$ from the effective potential.  However, this also implies
 the method is only valid for the large $k$ modes. This is not necessary in the discussion based on our new master equations. 
 
 Furthermore, by using  this new master equation, we show that effective metric or acoustic metric of the tensor perturbation equation defined by Reall in~\cite{Reall:2014pwa} can be generalized to the cases without a static condition.  In fact, we have get
 the effective metric $P^{MN}$ in Eq.(\ref{PMN5}).  With this effective metric in hand, we can study the hyperbolicity and causality of the tensor perturbation on 
all vacuum solutions of the theory.   We have given the explicit conditions that $P^{MN}$ is Lorentzian, i.e., $Q>0$ in Eq.(\ref{IEQ1}).  Under the assumption (\ref{static}), when condition (\ref{IEQ2}) and (\ref{IEQ3}) are satisfied, the causality is broken, and superluminary mode exists.

For each vacuum solution which can be written in the form of Eq.(\ref{metric}),  the exact hyperbolic condition of the tensor perturbation equations has been given. Among the black hole solutions, when $\tilde{M}>0$ and $1+4\tilde{\alpha}\tilde{\Lambda}>0$, $D=6$ or $n=4$ is very special because the hyperbolicity might be an issue outside the event horizon only in this case. We have found the analytic hyperbolic condition of the tensor perturbation equation on this background.  In the case $\Lambda=0$, only $K=1$ solution exists, and we find that hyperbolicity can be broken outside the event horizon, when the black hole is small enough, i.e., 
$$r_+< \sqrt{\tilde{\alpha}/3.0176} =1.4101 \sqrt{\alpha}\, . $$
This point has been noticed by Reall in~\cite{Reall:2014pwa}. Here, we have  found  the precise value of $r_+$ in the hyperbolic condition. In the case of $\Lambda<0$, $K=1$, we find the constraint by the hyperbolicity becomes tighter when $|\Lambda|$ increases.  
However, in the case of $\Lambda<0$, $K=0$, the situation is quite different: the radius of the black hole can be arbitrary small without breaking the hyperbolicity outside the event horizon if we turn down the abstract value of the cosmological constant.  For the positive $\Lambda$, the hyperbolic condition also provide constraint of the parameter $\alpha$, and detailed constraints have been depicted in the figures of the paper.  It should be noted here, for $\Lambda<0$ and  $K=-1$, in the case where mass parameter $\tilde{M}<0$, the  situation is complicated, and the hyperbolicity of the tensor perturbation might be broken for all $D\ge 5$. 
 Although the case of $1+4\tilde{\alpha}\tilde{\Lambda}<0$ is not so physically attractive,  for the completeness of the paper, we have also studied this situation with a very short discussion. For this background spacetime,  some wave equations exist when the spacetime dimension is greater than seven, i.e., $D>7$.
 
Nariai type solutions and dimensionally extended constant curvature black hole solutions are also considered in this paper, and the constraints by the hyperbolicity and causality on the tensor perturbation have been given explicitly. Our approach can also applied to dynamical spacetimes, and Vaidya spacetime have been investigated as an example.

We have discussed the case $m=2$, but from the Eq.(\ref{master1}) in Appendix A, we can also discuss the cases with $m=1$, and $m>2$. For instance, the case with $m=4$, which is interested for many people.  For example, our formulas can be used to the special Kaluza-Klein compactification in~\cite{Maeda:2006hj, Maeda:2006iw}, in which the topology of the spacetime is locally $\mathcal{M}^D\cong M^4\times N^{n}$ with a maximally symmetric space $N^n$ with negative sectional curvature.

On the other hand, the approach of using the effective metric to study the hyperbolicity and the causality can be applied to other  gravity theories, for example,  scalar-tensor gravity
and  $F(R)$ gravity theories.


\section*{Acknowledgement}
This work was supported in part by the National Natural Science Foundation of China with grants
No.11622543, No.12075232, No.11947301, and No.12047502.  This work is also supported by the 
Fundamental Research Funds for the Central Universities under Grant No: WK2030000036.



\appendix

\section*{Appendix.A}

\subsection{Background spacetimes}
In this appendix, we derive the perturbation equation for a general warped spacetime $\mathcal{M}^D\cong M^m\times N^n$ with metric
\begin{equation}
\label{metric1}
g_{MN}dx^Mdx^N=g_{ab}(y)dy^ady^b+r^2(y)\gamma_{ij}(z)dz^idz^j\, ,
\end{equation}
where coordinates $x^M=\{y^1\, ,\cdots y^m\, ;z^1\, ,\cdots\, ,z^n\}$.  The tuple $(M^m\, ,g_{ab})$ forms a $m-$dimension Lorentzian manifold, and $(N^n,\gamma_{ij})$ is an $n-$dimensional Riemann manifold. 
This Riemann manifold $(N^n,\gamma_{ij})$ is also assumed to be an Einstein manifold, i.e., the Ricci tensor is given by Eq.(\ref{Rij1}). 
As in section \ref{section2}, the metric compatible covariant derivatives associated with $g_{MN}$, $g_{ab}$, and $\gamma_{ij}$ are denoted by $\nabla_M$, $D_a$, and $\hat{D}_i$, respectively.

According to the metric (\ref{metric1}), we get the nontrivial components of Riemann tensor $R_{MNL}^{\quad\quad K}$ as follows:
\begin{eqnarray}
\label{riemann_tensor}
	{R}_{abc}^{\quad d}&=&{}^m\!{R}_{abc}^{\quad d}\, ,\nonumber\\
	R_{aib}^{\quad j}&=&-\frac{D_aD_br}{r}\delta_i^j\, ,\nonumber\\
	R_{ijk}^{\quad l}&=&\hat{R}_{ijk}^{\quad l}-(Dr)^2(\delta_j^l\gamma_{ki}-\delta_i^l\gamma_{kj})\, .
\end{eqnarray}
Here, ${}^m\!{R}_{abc}^{\quad d}$ and $\hat{R}_{ijk}^{\quad l}$ are the Riemann tensors of $(M^m\, ,r_{ab})$ and $(N^n\, ,\gamma_{ij})$, respectively, and $(Dr)^2=g^{ab}D_arD_br$. Utilizing Eqs.$(\ref{riemann_tensor})$, we obtain the nonvanished components of the Ricci tensor of the spacetime $(\mathcal{M}^D\, , g_{MN})$, which are given by
\begin{eqnarray}
	\label{ricci_tensor}
	R_{ab}&=&{}^m\!{R}_{ab}-n\frac{D_aD_br}{r}\, ,\nonumber\\
	R_{ij}&=&\left[-\frac{{}^m\!{\Box}r}{r}+(n-1)\frac{K-(Dr)^2}{r^2}\right]r^2\gamma_{ij}\, .
\end{eqnarray}
Then the scalar curvature of the spacetime has a form
\begin{equation}\label{scalar}
	R={}^m\!{R}-2n\frac{{}^m\!{\Box}r}{r}+n(n-1)\frac{K-(Dr)^2}{r^2}.
\end{equation}
In the above Eqs.(\ref{riemann_tensor})-(\ref{scalar}), ${}^m\!{R}$ and  ${}^m\!{R}_{ab}$ are the scalar curvature and the Ricci tensor of the manifold $(M^m\, ,g_{ab})$, respectively. The symbol ${}^2\!{\Box}=g^{ab}D_aD_b$ is the d'Alembertian in $(M^m\, ,g_{ab})$. With the metric (\ref{metric1}), any energy-momentum tensor $T_{MN}$ can be decomposed into $$T_{MN}=\mathrm{diag}\left\lbrace T_{ab}(y)\, ,r^2p(y)\gamma_{ij}\right\rbrace\, ,$$ in which both $T_{ab}$ and $p$  only depend on the coordinates $\left\lbrace y^a\right\rbrace$.

\subsection{General tensor perturbation equations}
For the background spacetime with the metric (\ref{metric1}), we discuss the tensor perturbation by setting 
\begin{eqnarray}
	\label{h}
		h_{ab}&=&0\, ,\quad\quad h_{ai}=0\, ,\nonumber\\
		\delta T_{ab}&=&0\, ,\quad\quad \delta T_{ai}=0\, ,
\end{eqnarray}
and the tensor $h_{ij}$ is transverse trace free i.e. $\hat{D}^ih_{ij}=0\, ,$ and
\begin{equation}
	h=g^{MN}h_{MN}=g^{ij}h_{ij}=\frac{\gamma^{ij}h_{ij}}{r^2}=r^2\gamma_{ij}h^{ij}=0\, .
\end{equation}
As a result, in Einstein-Gauss-Bonnet gravity theory, the nontrivial components of the perturbation equation (\ref{PE0}) are 
\begin{equation}
\label{Gij0}
	\delta G_{ij}+\Lambda h_{ij}+\alpha\delta H_{ij}=\kappa_{D}^2\delta T_{ij}\, .
\end{equation} 
To get the detailed forms of these equations, we have to calculate the  perturbation of the geometric quantities associated to the spacetime.   
Consider the perturbation of the metric $g_{MN}\to g_{MN}+h_{MN}$, one can  get 
\begin{eqnarray}
	\label{delta_riemann}
		\delta R_{MNL}^{\quad\quad P}&=&-\frac{1}{2}\Big[(\nabla_M\nabla_N-\nabla_N\nabla_M)h_L^{\ P}+(\nabla_M\nabla_L h_N^{\ P} \nonumber\\
		&-& \nabla_N\nabla_L h_M^{\ P})-(\nabla_M\nabla^P h_{NL}-\nabla_N\nabla^P h_{ML})\Big],
\end{eqnarray}
and $$\delta R_{MNLP}=g_{SP}\delta R_{MNL}^{\quad\quad S}+R_{MNL}^{\quad\quad S}h_{SP}\, .$$
By using Eq.$(\ref{h})$ and Eq. (\ref{delta_riemann}), we find that the perturbation of the Riemann tensor satisfies
\begin{equation}
\label{delta_riemann1}
	\delta R_{abcd} =0\, ,
\end{equation}
\begin{eqnarray}
	\delta R_{aibj}&=&\frac{2D_arD_br}{r^2}h_{ij}-\frac{1}{2}\left[\frac{D_arD_bh_{ij}}{r}+\frac{D_brD_ah_{ij}}{r}+r^2D_aD_b\left(\frac{h_{ij}}{r^2}\right)\right]-\frac{D_aD_br}{r}h_{ij}\, ,
	\label{delta_riemann2}
\end{eqnarray}
and
\begin{eqnarray}
\label{delta_riemann3}
\delta R_{ijkl}&=&\frac{r^2\gamma_{ml}}{2}(\hat{R}_{ijn}^{\quad m}h_k^{\ n}+\hat{R}_{ijk}^{\quad n}h_n^{\ m})-\frac{r^2\gamma_{ml}}{2}\left(\hat{D}_i\hat{D}_kh_j^{\ m}-\hat{D}_j\hat{D}_kh_i^{\ m}-\frac{1}{r^2}\hat{D}_i\hat{D}^mh_{jk}+\frac{1}{r^2}\hat{D}_j\hat{D}^mh_{ik}\right)\nonumber\\
&-&\frac{r^2\gamma_{ml}}{2}D^ar\left[r\gamma_{ik}D_ah_j^{\ m}-r\gamma_{jk}D_ah_i^{\ m}-\frac{\delta_i^m}{r}(D_ah_{jk})+\frac{\delta_j^m}{r}(D_ah_{ik})\right]+(Dr)^2(\gamma_{kj}h_{il}-\gamma_{ki}h_{jl})\, .
\end{eqnarray}
The perturbation of the Ricci tensor can be put into forms
\begin{eqnarray}
	\delta R_{ab}=\frac{2D_arD_br(\gamma^{ij}h_{ij})}{r^4}-\frac{\gamma^{ij}}{2r^3}(D_arD_bh_{ij}+D_brD_ah_{ij})-\frac{1}{2}D_aD_bh\, ,
\end{eqnarray}
and
\begin{eqnarray}
		\delta R_{ij}&=&-\frac{r^2}{2}\prescript{m}{}{\Box}\left(\frac{h_{ij}}{r^2}\right)-\frac{n}{2}\frac{D^ar}{r}D_ah_{ij}+\frac{1}{2r^2}\left(\hat{D}_i\hat{D}^kh_{jk}+\hat{D}_j\hat{D}^kh_{ik}\right)-\frac{1}{2r^2}\hat{\Delta}h_{ij}\nonumber\\
		&+&\left[(n-1)\frac{K}{r^2}+\frac{(Dr)^2}{r^2}-\frac{\prescript{m}{}{\Box r}}{r}\right]h_{ij}-\frac{1}{r^2}\hat{R}_{i\ j}^{\ k\ l}h_{kl}-\frac{1}{2}\hat{D}_i\hat{D}_jh-\frac{r}{2}D^arD_ah\gamma_{ij}\, .
\end{eqnarray}
By these, one can get the perturbation for the scalar curvature
\begin{eqnarray}
	\label{delta_scalar}
		\delta R&=&-\prescript{m}{}{\Box}h-\frac{1}{r^2}\hat{\Delta }h+\frac{1}{r^4}\hat{D}^i\hat{D}^jh_{ij}-\frac{1}{2}(n+2) \frac{D^ar}{r^3}\gamma^{ij}D_ah_{ij}\nonumber\\
		&&-\frac{n}{2r}D^arD_ah+\frac{(n+2)(Dr)^2}{r^2}h-\frac{K(n-1)h}{r^2}\, ,
\end{eqnarray}
where $\hat{\Delta}=\gamma^{ij}\hat{D}_i\hat{D}_j$ denotes the Laplace-Beltrami operator of $(N^n,\gamma_{ij})$.
Furthermore, by using Eqs.(\ref{delta_riemann1})-(\ref{delta_scalar}) and considering that the tensor $h_{ij}$ is transverse trace free, we obtain  the necessary terms to calculate  the perturbation of the Gauss-Bonnet tensor $H_{ij}$:
\begin{eqnarray}
\label{Prepare1}
		R\delta R_{ij}&=&\Bigg[\prescript{m}{}{R}-2n\frac{\prescript{m}{}{\Box}r}{r}+n(n-1)\frac{K-(Dr)^2}{r^2}\Bigg]\Bigg\{-\frac{r^2}{2}\prescript{m}{}{\Box}\left(\frac{h_{ij}}{r^2}\right)-\frac{n}{2}\frac{D^ar}{r}D_ah_{ij}\nonumber\\
		&&-\frac{1}{2r^2}\hat{\Delta}h_{ij}+\left[(n-1)\frac{K}{r^2}+\frac{(Dr)^2}{r^2}-\frac{\prescript{m}{}{\Box}r}{r}\right]h_{ij}-\frac{1}{r^2}\hat{R}_{i\ j}^{\ k\ l}h_{kl}\Bigg\}\, ,
\end{eqnarray}
\begin{eqnarray}
		R^M_{\ \ j}\delta R_{iM}&=&\left[-\frac{\prescript{m}{}{\Box}r}{r}+(n-1)\frac{K-(Dr)^2}{r^2}\right]\left\lbrace-\frac{r^2}{2}\prescript{m}{}{\Box}\left(\frac{h_{ij}}{r^2}\right)-\frac{n}{2}\frac{D^ar}{r}D_ah_{ij}-\frac{1}{2r^2}\hat{\Delta}h_{ij}\right.\nonumber\\
		&&\left.+\left[(n-1)\frac{K}{r^2}+\frac{(Dr)^2}{r^2}-\frac{\prescript{m}{}{\Box}r}{r}\right]h_{ij}-\frac{1}{r^2}\hat{R}_{i\ j}^{\ k\ l}h_{kl}\right\rbrace,
\end{eqnarray}
\begin{eqnarray}
R_{iM}\delta R^M_{\ \ j}&=&-\left[-\frac{\prescript{m}{}{\Box}r}{r}+(n-1)\frac{K-(Dr)^2}{r^2}\right]^2h_{ij}+\left[-\frac{\prescript{m}{}{\Box}r}{r}+(n-1)\frac{K-(Dr)^2}{r^2}\right]\left\lbrace-\frac{r^2}{2}\prescript{m}{}{\Box}\left(\frac{h_{ij}}{r^2}\right)\right.\nonumber\\
&&\left.-\frac{n}{2}\frac{D^ar}{r}D_ah_{ij}-\frac{1}{2r^2}\hat{\Delta}h_{ij}+\left[(n-1)\frac{K}{r^2}+\frac{(Dr)^2}{r^2}-\frac{\prescript{m}{}{\Box}r}{r}\right]h_{ij}-\frac{1}{r^2}\hat{R}_{i\ j}^{\ k\ l}h_{kl}\right\rbrace\, ,
\end{eqnarray}

\begin{eqnarray}
		R^{MN}\delta R_{iMjN}&=&\left(\prescript{m}{}{R}^{ab}-n\frac{D^aD^br}{r}\right)\Bigg\{\frac{2D_arD_br}{r^2}h_{ij}-\frac{1}{2}\left[\frac{D_arD_bh_{ij}}{r}+\frac{D_brD_ah_{ij}}{r}+r^2D_aD_b\left(\frac{h_{ij}}{r^2}\right)\right]\nonumber\\
		&&-\frac{D_aD_br}{r}h_{ij}\Bigg\}+\left[-\frac{\prescript{m}{}{\Box}r}{r}+(n-1)\frac{K-(Dr)^2}{r^2}\right]\Bigg\{\frac{K(n-1)h_{ij}}{r^2}-\frac{1}{2r^2}\hat{\Delta}h_{ij}\nonumber\\
		&&+\frac{r}{2}\gamma_{kj}D^arD_ah_i^{\ k}-\frac{n-1}{2r}D^arD_ah_{ij}+\frac{(Dr)^2}{r^2}h_{ij}\Bigg\}\, , 	
\end{eqnarray}

\begin{eqnarray}
R_{iMjN}\delta R^{MN}&=&-\frac{2}{r^2}\left[-\frac{\prescript{m}{}{\Box}r}{r}+(n-1)\frac{K-(Dr)^2}{r^2}\right]\left[\hat{R}_{i\ \ j}^{\ m\ n}h_{mn}+(Dr)^2h_{ij}\right]+\frac{1}{r^2}\hat{R}_{i\ j}^{\ k\ l}\left\lbrace-\frac{r^2}{2}\prescript{m}{}{\Box}\left(\frac{h_{kl}}{r^2}\right)\right.\nonumber\\
&&\left.-\frac{n}{2}\frac{D^ar}{r}D_ah_{kl}-\frac{1}{2r^2}\hat{\Delta}h_{kl}+\left[(n-1)\frac{K}{r^2}+\frac{(Dr)^2}{r^2}-\frac{\prescript{m}{}{\Box}r}{r}\right]h_{kl}-\frac{1}{r^2}\hat{R}_{k\ \ l}^{\ m\ n}h_{mn}\right\rbrace\nonumber\\
&&+\frac{(Dr)^2}{r^2}\Bigg\{-\frac{r^2}{2}\prescript{m}{}{\Box}\left(\frac{h_{ij}}{r^2}\right)-\frac{n}{2}\frac{D^ar}{r}D_ah_{ij}-\frac{1}{2r^2}\hat{\Delta}h_{ij}+\Big[(n-1)\frac{K}{r^2}+\frac{(Dr)^2}{r^2}\nonumber\\
&&-\frac{\prescript{m}{}{\Box}r}{r}\Big]h_{ij}-\frac{1}{r^2}\hat{R}_{i\ j}^{\ k\ l}h_{kl}\Bigg\}\, ,
\end{eqnarray}

\begin{eqnarray}
		R_i^{\ MNP}\delta R_{jMNP}&=&-\frac{2D^aD^br}{r}\left\lbrace\frac{2D_arD_br}{r^2}h_{ij}-\frac{1}{2}\left[\frac{D_arD_bh_{ij}}{r}+\frac{D_brD_ah_{ij}}{r}+r^2D_aD_b\left(\frac{h_{ij}}{r^2}\right)\right]-\frac{D_aD_br}{r}h_{ij}\right\rbrace\nonumber\\
		&&+\frac{1}{r^4}\hat{R}_i^{\ mkl}\Bigg\{\frac{r^2}{2}\gamma_{pl}\left(\hat{R}_{jmn}^{\quad\ p}h_k^{\ n}+\hat{R}_{jmk}^{\quad\ n}h_n^{\ p}\right)-\frac{r^2\gamma_{pl}}{2}\Big[\hat{D}_j\hat{D}_kh_m^{\ p}-\hat{D}_m\hat{D}_kh_j^{\ p}-\frac{1}{r^2}\hat{D}_j\hat{D}^ph_{mk}\nonumber\\
		&& +\frac{1}{r^2}\hat{D}_m\hat{D}^ph_{jk}\Big]\Bigg\}-\frac{r^2\gamma_{pl}}{2}D^ar\Big[r\gamma_{jk}(D_ah_m^{\ p})-r\gamma_{mk}(D_ah_j^{\ p})-\frac{\delta_j^p}{r}(D_ah_{mk})+\frac{\delta_m^p}{r}(D_ah_{jk})\nonumber\\
		&&+(Dr)^2(\gamma_{km}h_{jl}-\gamma_{kj}h_{ml})\Big]-\frac{2(Dr)^2}{r^4}K(n-1)h_{ij}+\frac{(Dr)^2}{r^4}\hat{\Delta}h_{ij}+\frac{(n-2)[(Dr)^2]^2}{r^4}h_{ij}\nonumber\\
		&&+\frac{(Dr)^2D^ar}{2r^2}(n-2)\left(r\gamma_{mi}D_ah_j^{\ m}+\frac{D_ah_{ij}}{r}\right)\, ,
\end{eqnarray}
\begin{eqnarray}
		R_{jMNP}\delta R_i^{\ MNP}&=&-\frac{2D^aD^br}{r}\Bigg\{\frac{2D_arD_br}{r^2}h_{ij}-\frac{1}{2}\left[\frac{D_arD_bh_{ij}}{r}+\frac{D_brD_ah_{ij}}{r}+r^2D_aD_b\left(\frac{h_{ij}}{r^2}\right)\right]\Bigg\}\nonumber\\
		&&+\frac{1}{r^4}\hat{R}_j^{\ mkl}\Bigg\{\frac{r^2}{2}\gamma_{pl}(\hat{R}_{imn}^{\quad\ p}h_k^{\ n}+\hat{R}_{imk}^{\quad\ n}h_n^{\ p})-\frac{r^2\gamma_{pl}}{2}\Big[\hat{D}_i\hat{D}_kh_m^{\ p}-\hat{D}_m\hat{D}_kh_i^{\ p}-\frac{1}{r^2}\hat{D}_i\hat{D}^ph_{mk}\nonumber\\
		&& +\frac{1}{r^2}\hat{D}_m\hat{D}^ph_{ik}\Big]\Bigg\}-\frac{r^2\gamma_{pl}}{2}D^ar\Big[r\gamma_{ik}(D_ah_m^{\ p})-r\gamma_{mk}(D_ah_i^{\ p})-\frac{\delta_i^p}{r}(D_ah_{mk})\nonumber\\
		&&+\frac{\delta_m^p}{r}(D_ah_{ik})+(Dr)^2(\gamma_{km}h_{il}-\gamma_{ki}h_{ml})\Big] -\frac{2(Dr)^2}{r^4}K(n-1)h_{ij}+\frac{(Dr)^2}{r^4}\hat{\Delta}h_{ij}\nonumber\\
		&&+\frac{(Dr)^2D^ar}{2r^2}(n-2)\left(r\gamma_{mj}D_ah_i^{\ m}+\frac{D_ah_{ij}}{r}\right)+\frac{(4-n)[(Dr)^2]^2h_{ij}}{r^4}\nonumber\\
		&&-\hat{R}_{jmnp}\hat{R}_{istu}\left(h^{ms}\gamma^{nt}\gamma^{pu}+h^{nt}\gamma^{ms}\gamma^{pu}+h^{pu}\gamma^{ms}\gamma^{nt}\right)\nonumber\\
		&&+\frac{4(Dr)^2}{r^4}\left[2\hat{R}_{i\ j}^{\ s\ t}h_{st}+K(n-1)h_{ij}\right]\, ,
\end{eqnarray}
and
\begin{equation}
\label{Prepare2}
	\delta L_{GB} =-\frac{2\hat{R}^{ijkl}\hat{R}_{ijk}^{\quad n}h_{nl}}{r^6}-\frac{\hat{R}^{ijkl}}{r^6}\left(\hat{D}_i\hat{D}_kh_{jl}-\hat{D}_j\hat{D}_kh_{il}-\hat{D}_i\hat{D}_lh_{jk}+\hat{D}_j\hat{D}_lh_{ik}\right)\, .
\end{equation}
From Eq.(\ref{GB_tensor}),  we have
\begin{eqnarray}
\label{delta_Hij}
		\delta H_{ij}&=&2(R\delta R_{ij}+R_{ij}\delta R)-4(R^M_{\ \ j}\delta R_{iM}+R_{iM}\delta R^M_{\ \ j})-4(R^{MN}\delta R_{iMjN}+R_{iMjN}\delta R^{MN})\nonumber\\
		&&+2(R_i^{\ MNP}\delta R_{jMNP}+R_{jMNP}\delta R_i^{\ MNP})-\frac{1}{2}h_{ij}L_{GB}-\frac{1}{2}g_{ij}\delta L_{GB}\, .
\end{eqnarray}
For the Einstein manifold, the relation between the  Weyl tensor and  the Riemann tensor is given by 
\begin{equation}
\label{Weyl11}
\hat{C}_{ijkl}=\hat{R}_{ijkl}-K(\gamma_{ik}\gamma_{lj}-\gamma_{il}\gamma_{kj})\, .
\end{equation}
Substituting Eqs.(\ref{Prepare1})-(\ref{Prepare2}), and Eq.$(\ref{Weyl11})$ into Eq.(\ref{delta_Hij}), after lengthy calculation, we finally obtain the exact formula of $\delta H_{ij}$:
\begin{eqnarray}
\label{deltaHij1}
\frac{2 \delta H_{ij}}{r^2}&=&4 \Bigg\{{}^m\!{G}^{ab}- (n-2)\frac{D^aD^br}{r}-\left[\frac{1}{2}(n-2)(n-3)\frac{K-(Dr)^2}{r^2}-(n-2)\frac{\prescript{m}{}{\Box}r}{r}\right]g^{ab}\Bigg\}D_aD_b\left(\frac{h_{ij}}{r^2}\right)\nonumber\\
&+&8\Bigg\{  {}^m\!{G}^{ab}-(n-2)\frac{D^aD^br}{r}-\frac{1}{4}(n-2) \Bigg[{}^m\!{R}- 2(n-1)\frac{\prescript{m}{}{\Box}r}{r}+(n-2)(n-3)\frac{K-(Dr)^2}{r^2}\Bigg]g^{ab}\Bigg\}\nonumber\\
&& \times\frac{D_br}{r}D_a\left(\frac{h_{ij}}{r^2}\right)+2\left[- \frac{{}^m\!{R}}{r^2}+\frac{2(n-3)\prescript{m}{}{\Box}r}{r^3}-(n-3)(n-4)\frac{K-(Dr)^2}{r^4}\right]\hat{\Delta}\left(\frac{h_{ij}}{r^2}\right)\nonumber\\
&+&4 \Bigg\{ -{}^m\!{R}\cdot \frac{{}^m\!{\Box}r}{r}+2(n-1)\Big(\frac{\prescript{m}{}{\Box}r}{r }\Big)^2+n ~ {}^m\! {R}\cdot \frac{K}{r^2}-(n-1)~{}^m\! {R}\cdot \frac{(Dr)^2}{r^2} - n(3n-7)\frac{K}{r^2}\cdot \frac{\prescript{m}{}{\Box}r}{r}\nonumber\\
&&+3(n-1)(n-2)(n-3) \frac{(Dr)^2}{r^2}\cdot \frac{\prescript{m}{}{\Box}r}{r}+(n-1)(n-2)(n-3) \Big[\frac{(Dr)^2}{r^2}\Big]^2\nonumber\\
&&+ (n-3)(n^2-2n-2) \Big(\frac{K}{r^2}\Big)^2- n(n-3)(2n-5)\frac{ K}{r^2}\cdot \frac{ (Dr)^2}{r^2}\nonumber\\
&&2 \left[\prescript{m}{}{R}^{ab}-(n-1)\frac{D^aD^br}{r}\right]\frac{D_aD_br}{r}-\frac{1}{4}L_{GB}\Bigg\}\left(\frac{h_{ij}}{r^2}\right)+\frac{2W_{ij}}{r^2} \, .
\end{eqnarray}
Here
\begin{eqnarray}
L_{GB}&=& {}^m\! L_{GB} + 8n~{}^m\! G^{ab}\frac{D_aD_b r}{r} - 4n(n-1)(n-2) \frac{{}^m\! \Box r}{r}\cdot \frac{K-(Dr)^2}{r^2}\nonumber\\
&+&n(n-1)(n-2)(n-3) \Bigg[ \frac{K-(Dr)^2}{r^2}\Bigg]^2 -4n(n-1)\frac{(D_{a}D_br)(D^aD^br)}{r^2}\nonumber\\
&+&2n(n-1) ~ {}^m\! R\cdot \frac{K-(Dr)^2}{r^2} + 4n(n-1) \Big(\frac{{}^m\! \Box r}{r}\Big)^2 + \frac{\hat{C}_{ijkl}\hat{C}^{ijkl}}{r^4}\, ,
\end{eqnarray}
where ${}^m\! L_{GB}$ is the Gauss-Bonnet term in the Lorentizan manifold $(M^m\, ,g_{ab})$. 
The  $W_{ij}$ in Eq.(\ref{deltaHij1}) contains the terms about the Weyl tensor $\hat{C}_{ijkl}$, and it can be expressed as
\begin{eqnarray}
\label{Wij1}
W_{ij}&=&2\hat{C}_{i\ j}^{\ k\ l}\prescript{m}{}{\Box}\left(\frac{h_{kl}}{r^2}\right)+\frac{2n-4}{r}\hat{C}_{i\ j}^{\ k\ l}D^arD_a\left(\frac{h_{kl}}{r^2}\right)+\frac{2}{r^2}\hat{C}_{i\ j}^{\ k\ l}\hat{\Delta}\left(\frac{h_{kl}}{r^2}\right)\nonumber\\
&-&\frac{\hat{C}_i^{\ mkl}}{r^4}\left[\hat{D}_j\hat{D}_kh_{ml}-\hat{D}_m\hat{D}_kh_{jl}-\hat{D}_j\hat{D}_lh_{mk}+\hat{D}_m\hat{D}_lh_{jk}\right]\nonumber\\
&-&\frac{\hat{C}_j^{\ mkl}}{r^4}\left[\hat{D}_i\hat{D}_kh_{ml}-\hat{D}_m\hat{D}_kh_{il}-\hat{D}_i\hat{D}_lh_{mk}+\hat{D}_m\hat{D}_lh_{ik}\right]\nonumber\\
&+&\frac{\hat{C}^{pqkl}}{2r^4}\left[\hat{D}_p\hat{D}_kh_{ql}-\hat{D}_q\hat{D}_kh_{pl}-\hat{D}_p\hat{D}_lh_{qk}+\hat{D}_q\hat{D}_lh_{pk}\right]\gamma_{ij}\nonumber\\
&+&\frac{4}{r^4}\hat{C}_{i\ j}^{\ k\ l}\hat{C}_{k\ l}^{\ m\ n}h_{mn}+\frac{2}{r^4}\hat{C}_{j\ n}^{\ k\ m}\hat{C}_{m\ i}^{\ n\ l}h_{kl}+\frac{\gamma_{ij}}{r^4}\hat{C}^{pqkl}\hat{C}_{pqk}{}^n h_{nl}\nonumber\\
&+&2 \Big[-{}^m\!{R}+2(n-3) \frac{\prescript{m}{}{\Box}r}{r}-(n^2-7n+16)\frac{K}{r^2}\nonumber\\
&&+(n-3)(n-4)\frac{(Dr)^2}{r^2}\Big]\hat{C}_{i\ j}^{\ k\ l}\Big(\frac{h_{kl}}{r^2}\Big)\, .
\end{eqnarray}
At the same time, we can also get the perturbation of the Einstein tensor $\delta G_{ij}$
\begin{eqnarray}
\label{Gij1}
\frac{2 \delta G_{ij}}{r^2}&=&- \prescript{m}{}{\Box}\left(\frac{h_{ij}}{r^2}\right)-n\frac{D^ar}{r }D_a\left(\frac{h_{ij}}{r^2}\right)+\frac{\hat{\Delta}_L}{r^2}\left(\frac{h_{ij}}{r^2}\right)-\Bigg[ \prescript{m}{}{R}-2(n-1)\frac{\prescript{m}{}{\Box}r}{r}\nonumber\\
&&+n(n-1)\frac{K}{r^2}-(n-1)(n-2)\frac{(Dr)^2}{r^2}\Bigg]\left(\frac{h_{ij}}{r^2}\right)\, ,  	
\end{eqnarray}
where $\hat{\Delta}_L$ is the Lichnerowicz operator acting on the symmetric rank-2 tensor on $(N^n,\gamma_{ij})$. The relation between this operator and usual Laplace operator is given by the following formula,
\begin{equation}
\label{deltaL1}
\hat{\Delta}_Ls_{ij}=-\hat{\Delta}s_{ij}+\hat{R}_i^{\ k}s_{kj}+\hat{R}_j^{\ k}s_{ik}-2\hat{R}_{i\ j}^{\ k\ l}s_{kl},
\end{equation}
where $s_{ij}$ is an arbitrary symmetric tensor field tensor on $(N^n,\gamma_{ij})$. When $(N^n,\gamma_{ij})$ is maximal symmetry manifold, it should be noted here that $$\hat{\Delta}_Ls_{ij}=(-\hat{\Delta}+2nK)s_{ij}\, .$$ 
From Eqs.(\ref{deltaHij1}), (\ref{Wij1}), (\ref{Gij1}), and (\ref{deltaL1}),  we find that Eq.(\ref{Gij0}) becomes
\begin{eqnarray}
\label{master1}
\Big(P^{ab}{}_{ij}{}^{kl} D_aD_b + P^{mn}{}_{ij}{}^{kl} \hat{D}_m\hat{D}_n + P^{a}{}_{ij}{}^{kl} D_a + V_{ij}{}^{kl}\Big)\Big(\frac{h_{kl}}{r^2}\Big) = -\frac{2\kappa_D^2}{r^2}\delta T_{ij}\, ,
\end{eqnarray}
where
\begin{eqnarray}
P^{ab}{}_{ij}{}^{kl} = P^{ab} \delta_i{}^k\delta_j{}^l -\frac{4\alpha}{r^2}g^{ab} \hat{C}_i{}^k{}_j{}^l \, ,
\end{eqnarray}
\begin{eqnarray}
P^{a}{}_{ij}{}^{kl} = P^{a} \delta_i{}^k\delta_j{}^l -4\alpha(n-2)\frac{D^ar}{r} \frac{\hat{C}_i{}^k{}_j{}^l}{r^2}  \, ,
\end{eqnarray}
\begin{eqnarray}
P^{mn}{}_{ij}{}^{kl} = P^{mn} \delta_i{}^k\delta_j{}^l + \frac{4\alpha}{r^2}\big(\hat{C}_j{}^{knl} \delta_i{}^m+\hat{C}_i{}^{knl} \delta_j{}^m+\hat{C}_j{}^{mln} \delta_i{}^k +\hat{C}_i{}^{mln} \delta_j{}^k- \hat{C}^{mknl}\gamma_{ij} - \hat{C}_i{}^k{}_j{}^l\gamma^{mn} \big) \, ,
\end{eqnarray}
\begin{eqnarray}
V_{ij}{}^{kl} &=& V \delta_i{}^k\delta_j{}^l + \frac{2 \hat{C}_i{}^k{}_j{}^l}{r^2} +\alpha \Bigg\{ 4\Bigg[{}^m\!{R} - 2(n-3)\frac{{}^m\!\Box r}{r}  + (n^2 - 7n + 16)\frac{K}{r^2} - (n-3)(n-4)\frac{(Dr)^2}{r^2}\Bigg] \frac{ \hat{C}_i{}^k{}_j{}^l}{r^2}\nonumber\\
&&-\frac{8}{r^4} \hat{C}_{imjn}\hat{C}^{mknl} + \frac{4}{r^4} \hat{C}_{mn j}{}^k{}\hat{C}^{mn}{}_{i}{}^l -\frac{2}{r^4}\hat{C}^{mnpl}\hat{C}_{mnp}{}^k\gamma_{ij }+\frac{\hat{C}^{mnpq}\hat{C}_{mnpq}}{r^4}\delta_i{}^k\delta_j{}^l \Bigg\} \, ,
\end{eqnarray}
In the above equations
\begin{eqnarray}
P^{ab}= g^{ab} + 2(n-2) \alpha\left\lbrace 2\frac{D^aD^br}{r}+\left[(n-3)\frac{K-(Dr)^2}{r^2}-2\frac{\prescript{m}{}{\Box}r}{r}\right]g^{ab}\right\rbrace-4\alpha\cdot {}^m\! G^{ab}\, ,
\end{eqnarray}
\begin{eqnarray}
P^{mn}=\Bigg\{1
+2 \alpha\left[ {}^m\!{R}- \frac{2(n-3) {}^m\!{\Box}r}{r}+ (n-3)(n-4)\frac{K-(Dr)^2}{r^2}\right] \Bigg\}\frac{\gamma^{mn }}{r^2}\, ,
\end{eqnarray}
\begin{eqnarray}
P^{a}&=&n\frac{D^ar}{r} +  2(n-2) \alpha \Bigg\{4\frac{D^aD^br}{r}+\Big[ {}^m\!{R}
-2(n-1)\frac{ {}^m\!{\Box}r}{r}\nonumber\\ 
&&+(n-2) (n-3)\frac{K-(Dr)^2}{r^2}\Big]g^{ab}\Bigg\}\frac{D_br}{r}- 8\alpha\cdot {}^m\! G^{ab}\frac{D_br}{r}\, ,
\end{eqnarray}
and
\begin{eqnarray}
V&=&{}^m\!{R}-2(n-1)\frac{{}^m\! {\Box}r}{r}+\frac{n(n-3)K}{r^2}-\frac{(n-1)(n-2)(Dr)^2}{r^2}-\Lambda\nonumber\\
&+&\alpha\Bigg\{{}^m\! L_{GB}  + 8(n-1)\cdot {}^m\! G^{ab}\frac{D_aD_br}{r} -4(n-1)(n-2)\frac{(D^aD^br)(D_aD_br)}{r^2}\nonumber\\
&+& 4(n-1)(n-2)\left(\frac{{}^m\! {\Box}r}{r}\right)^2+2n(n-3)\frac{K\cdot {}^m\! R}{r^2}-2(n-1)(n-2)\frac{(Dr)^2\cdot {}^m\! {R}}{r^2}\nonumber\\
&-&4n(n-3)^2\frac{K\cdot {}^m\! {\Box}r}{r^3}+4(n-1)(n-2)(n-3)\frac{(Dr)^2\cdot{}^m\! {\Box}r}{r^3}\nonumber\\
&-&2n(n-3)^2(n-4)\frac{K\cdot(Dr)^2}{r^4}+(n-3)(n-4)(n^2-3n-2)\frac{K^2}{r^4}\nonumber\\
&+&(n-1)(n-2)(n-3)(n-4)\left[\frac{(Dr)^2}{r^2}\right]^2\Bigg{\}}\, .
\end{eqnarray}
Eq.(\ref{master1}) is the most general master equation of tensor type for the warped spacetime with the metric (\ref{metric1}).   In general, the components of $h_{ij}=h^{\mathrm{TT}}_{ij}$ are coupled to each other
if $(N\, ,\gamma_{ij})$ is not maximally symmetric. 
If we restrict to the case with $m=2$, we have ${}^2\!G_{ab}=0$, ${}^2\!L_{GB}=0$, and the above equations reduce to Eqs.(\ref{Pab}), (\ref{PijQ1}), (\ref{Pa123}), (\ref{V123}) in Sec.\ref{section3}.  If we further restrict to
the case that  $(N\, ,\gamma_{ij})$ is  maximally symmetric, Eq.(\ref{master1}) reduces to Eq.(\ref{PE1}) in Sec.\ref{section3}.

\section*{Appendix.B}
In two-dimensional Lorentz manifold, the volume element $\epsilon_{ab}$ and the metric $g_{ab}$ can be related by
\begin{equation}
\epsilon_{ac}\epsilon_{bd}=g_{ad}g_{cb}-g_{ab}g_{cd}.
\end{equation}
Hence, the determinant of $P^{ab}$ can be expressed as
\begin{equation}
P=\frac{1}{2}\epsilon_{ac}\epsilon_{bd}P^{ab}P^{cd}=\frac{1}{2}P_{ab}P^{ab}-\frac{1}{2}\left[\mathrm{Tr}( P)\right]^2\, ,
\end{equation}
where $\mathrm{Tr}(P)=g_{ab}P^{ab}$ is the trace of the tensor $P^{ab}$. Simple calculations show
\begin{equation}
\mathrm{Tr}(P)=2+\alpha\left[4(n-2)(n-3)\frac{K-(Dr)^2}{r^2}-4(n-2)\frac{\prescript{2}{}{\Box}r}{r}\right]\, ,
\end{equation}
and
\begin{eqnarray}
P_{ab}P^{ab}&=&2+2\alpha\left[4(n-2)(n-3)\frac{K-(Dr)^2}{r^2}-4(n-2)\frac{\prescript{2}{}{\Box}r}{r}\right]\nonumber\\
&&+\alpha^2\Bigg{\{}8(n-2)\frac{\prescript{2}{}{\Box}r}{r}\left[2(n-2)(n-3)\frac{K-(Dr)^2}{r^2}-4(n-2)\frac{\prescript{2}{}{\Box}r}{r}\right]\nonumber\\
&&+2\left[2(n-2)(n-3)\frac{K-(Dr)^2}{r^2}-4(n-2)\frac{\prescript{2}{}{\Box}r}{r}\right]^2\nonumber\\
&&+16(n-2)^2\frac{(D_aD_br)(D^aD^br)}{r^2}\Bigg{\}}\, .
\end{eqnarray}
By using expressions of $\text{Tr}(P)$ and $P_{ab}P^{ab}$ in the above equations, we arrive at
\begin{eqnarray}
\label{PAppendix}
P&=&-1-\alpha\left[4(n-2)(n-3)\frac{K-(Dr)^2}{r^2}-4(n-2)\frac{\prescript{2}{}{\Box}r}{r}\right]\nonumber \\
&&+\alpha^2\Bigg{\{}8(n-2)^2\frac{(D_aD_br)(D^aD^br)}{r^2}-8(n-2)^2\left(\frac{\prescript{2}{}{\Box}r}{r}\right)^2 \nonumber\\
&&+8(n-2)^2(n-3)\frac{\prescript{2}{}{\Box}r[K-(Dr)^2]}{r^3}\nonumber\\
&&-4(n-2)^2(n-3)^2\frac{[K-(Dr)^2]^2}{r^4}\Bigg{\}}. 
\end{eqnarray}
Consider Eq.(\ref{EOM_2}), we obtain
\begin{align}
	D_aD_br- \frac{1}{2}\prescript{2}{}{\Box}rg_{ab}=  \frac{r}{n}\kappa_{D}^2\left(\frac{1}{2}g^{cd}T_{cd}g_{ab}-T_{ab}\right)\cdot \Bigg[1+2\alpha(n-1)(n-2)\frac{K-(Dr)^2}{r^2}\Bigg]^{-1}\, .
\end{align}
and
\begin{eqnarray}
(D_aD_br)(D^aD^br)- \frac{1}{2} (\prescript{2}{}{\Box}r)^2 &=& \frac{r^2}{n^2}\kappa_{D}^4\left(\frac{1}{2}g^{cd}T_{cd}g_{ab}-T_{ab}\right)\left(\frac{1}{2}g^{ef}T_{ef}g^{ab}-T^{ab}\right)\nonumber\\
&&\times \Bigg[1+2\alpha(n-1)(n-2)\frac{K-(Dr)^2}{r^2}\Bigg]^{-2}
\label{DDrDDr}.
\end{eqnarray}
We can choose null frame $\{ n^a\, ,\ell^a\}$ such that
$$g_{ab}=-\ell_an_b-n_a\ell_b\, ,$$ 
where
$$\ell_a\ell^a=0=n_an^a\, ,\quad \ell_an^a=-1\, .$$
$T_{ab}$ can be expressed as 
\begin{equation}
T_{ab}=T_{\ell\ell}n_an_b+T_{nn}\ell_a\ell_b+\frac{1}{2}Tg_{ab}\, ,
\label{T_{ab}}
\end{equation}
where $$T_{\ell\ell}=T_{ab}\ell^a\ell^b\, ,\qquad T_{nn}=T_{ab}n^an^b\, ,$$ 
and 
$$T=g^{ab}T_{ab}=-2n^a\ell^bT_{ab}\, .$$
Finally, by using Eq.(\ref{DDrDDr}) and Eq.(\ref{T_{ab}}), one gets Eq.(\ref{P_nomal}).

\end{document}